\documentclass[utf8]{frontiersSCNS} 

\usepackage{url,hyperref,lineno,microtype,subcaption,ulem}
\usepackage[onehalfspacing]{setspace}
\usepackage[colorinlistoftodos]{todonotes}

\newcommand{\gn}{\color{magenta}}

\def\comma{{\gn,}~}

\def\apjl{The Astrophysical Journal Letters}
\def\apjs{The Astrophysical Journal Supplement}
\def\aap{Astronomy and Astrophysics}

\def\keyFont{\fontsize{8}{11}\helveticabold }
\def\firstAuthorLast{Kuroda {et~al.}} 
\def\Authors{Natsuha Kuroda\,$^{1,2*}$, Gregory D. Fleishman\,$^{3}$, Dale E. Gary\,$^{3}$, Gelu M. Nita\,$^{3}$, Bin Chen\,$^{3}$, Sijie Yu\,$^{3}$}
\def\Address{$^{1}$University Corporation for Atmospheric Research, Boulder, CO \\
$^{2}$Naval Research Laboratory, Washington DC \\
$^{3}$New Jersey Institute of Technology, Newark, New Jersey}
\def\corrAuthor{Natsuha Kuroda}

\def\corrEmail{nkuroda@ucar.edu}

\begin{document}
\onecolumn
\firstpage{1}

\title[Evolution of Flare-accelerated Electrons]{ Evolution of Flare-accelerated Electrons Quantified  by Spatially Resolved  Analysis}

\author[\firstAuthorLast ]{\Authors} 
\address{} 
\correspondance{} 

\extraAuth{}

\maketitle

\begin{abstract}

\section{}

Nonthermal electrons accelerated in solar flares produce electromagnetic emission in two distinct, highly complementary domains---hard X-rays (HXRs) and microwaves (MWs). This paper reports MW imaging spectroscopy observations from the Expanded Owens Valley Solar Array of an M1.2 flare that occurred on 2017 September 9, from which we deduce evolving coronal parameter maps. We analyze these data jointly with the complementary Reuven Ramaty High-Energy Solar Spectroscopic Imager HXR data to reveal the spatially-resolved evolution of the nonthermal electrons in the flaring volume. We find that the high-energy portion of the nonthermal electron distribution, responsible for the MW emission, displays a much more prominent evolution (in the form of strong spectral hardening) than the low-energy portion, responsible for the HXR emission. We show that the revealed trends are consistent with a single electron population evolving according to a simplified trap-plus-precipitation model with sustained injection/acceleration of nonthermal electrons, which produces a double-powerlaw with steadily increasing break energy.

\tiny
 \keyFont{ \section{Keywords:} solar flares, microwave, imaging spectroscopy, nonthermal electrons, numerical modeling, X-ray, corona, evolution} 
\end{abstract}

\section{Introduction}
\label{sec:intro}

Solar flares are the manifestations of free magnetic energy conversion to other forms---thermal, nonthermal, and kinetic. Often, a large fraction of this energy goes into acceleration of ambient charged particles \citep{1971SoPh...17..412L,2012ApJ...759...71E,2016ApJ...832...27A}, making the nonthermal particles dynamically and energetically important.
Probing the accelerated electrons may be done by exploiting the nonthermal emissions they produce---the hard X-ray (HXR) and microwave (MW) spatial, spectral, and temporal signatures. 
HXRs are produced by bremsstrahlung from dense regions, as a signature of either the footpoint bombardment by the  electron beams  \citep[e.g.,][]{1981ApJ...246L.155H} or dense coronal regions, which might be the particle acceleration region itself \citep{1994Natur.371..495M,2010ApJ...714.1108K,2014ApJ...780..107K}. The MWs are dominated by the gyrosynchrotron (GS) emission due to nonthermal electrons gyrating in the coronal magnetic field with a contribution from free-free emission. 
As a result of these distinct emission mechanisms, even a single population of nonthermal electrons distributed over a single (but possibly magnetically-asymmetric) flaring loop yields spatially-displaced HXR and MW emissions \citep[e.g.,][]{2016ApJ...816...62F}.
While most of the HXR spectrum is formed by nonthermal electrons with energies from a few to a (few) hundred keV, the spectrum of the GS-emitting electrons may extend to much higher energies including the MeV range \citep{1986SoPh..105...73N,1994ApJS...90..599K}.   In the complex magnetic topology of a solar flare, the flare-accelerated electrons tend to fill out any magnetic flux tube to which they have access. The  magnetic-field-dominated GS emission can be strong even from those spatial locations that are HXR-faint due to their low ambient density.
Indeed, \citet{2018ApJ...867...81F} found that MW low frequency sources, indicative of low magnetic field high in the corona, are typically much larger than the HXR sources. Therefore, the HXR and MW data offer complementary information on both the energy and spatial distributions of the nonthermal electrons. This implies that both spectrally and spatially resolved data are needed to probe the nonthermal electrons most comprehensively.

Compared with the HXRs \citep{2011SSRv..159..225W}, the spatial distribution of MW-emitting electrons is often much larger (and richer in complexity), so MW emission is well suited to quantify the accelerated electrons in space.  \citet{2018ApJ...867...84G} studied the nonthermal electrons in a flare-jet configuration  and found an equipartition of the nonthermal energies between populations in the closed and open magnetic flux tubes.  
Closed flaring flux tubes can represent rather large reservoirs of high-energy electrons located either nearby
\citep{2018ApJ...852...32K} or far away from \citep{2017ApJ...845..135F} the main flare acceleration sites, possibly  providing the  seed population for solar energetic particles (SEPs).
\citet{Fl_etal_2011,2013ApJ...768..190F,2016ApJ...826...38F}  probed the acceleration sites using MW observations and  concluded that the acceleration regime was consistent with stochastic acceleration, while 
\citet{2018ApJ...859...17F} extended in time the studies of  \citet{2016ApJ...816...62F} and \citet{2018ApJ...852...32K} to quantify the acceleration and transport of the nonthermal electrons in the 3D domain. 
In all of these studies, broadband MW spectroscopy and imaging have been crucial. 

Until recently, a critical element was missing from the observations---the ability to make high-fidelity MW images at many frequencies from which to obtain spatially-resolved spectra. This ability has become available with the Expanded Owens Valley Solar Array \citep[EOVSA;][]{2018ApJ...863...83G,2016JAI.....541009N}. This solar-dedicated radio interferometer can image flares anywhere on the solar disk at hundreds of frequencies spread over 1--18 GHz at 1~s cadence. The spatially-resolved spectrum from each pixel in the high-resolution images obtained with EOVSA can be forward-fit (Nita et al. 2020, in preparation) with a ``cost function'' that accounts for GS and free--free radio emission and absorption \citep{Fleishman278}.  As a result, one can now simultaneously obtain all relevant physical parameters over the entire source region at the desired cadence down to 1~s \citep[][]{2009ApJ...698L.183F,2013SoPh..288..549G, Fleishman278}. This novel and unique methodology allows the quantitative study of the spatial distribution and the temporal evolution of the magnetic field and the plasma in the corona in much greater detail than was previously possible.

Since the start of full operations in April 2017, EOVSA has recorded MW imaging spectroscopic observations of dozens of flares in all sizes, including some of the largest flares in Solar Cycle 24, which occurred during the 2017 September period \citep{2018ApJ...863...83G}. Previous papers using EOVSA imaging data have all focused on the well-observed 2017 Sep 10 flare (the second largest X-class flare of solar cycle \# 24) \citep{2018ApJ...863...83G,Fleishman278}.  This paper reports observations of a second flare observed during the same 2017 September period, a mid-sized M1.2 flare that occurred on 2017 Sep 9. 
We employ the MW forward-fitting technique, augmented by observations in HXRs and extreme ultra-violet (EUV) available from the \textit{Reuven Ramaty High-Energy Solar Spectroscopic Imager} \citep[\textit{RHESSI};][]{2002SoPh..210....3L} and the \textit{Atmospheric Imaging Assembly} \citep[\textit{AIA;}][]{2012SoPh..275...17L} on the \textit{Solar Dynamics Observervatory} (\textit{SDO}), respectively. In paricular,
we focus on the comparison of the electrons emitting MW in the corona and those emitting nonthermal, thick-target HXRs in the lower atmosphere.


\section{Multi-wavelengths Observation}
\label{sec:observation}
The M1.2 flare (SOL2017-09-09T22:04) started at around 22:04 UT and peaked around 23:53 UT in the 1.0--8.0~\AA~channel of the \textit{Geostationary Operational Environmental Satellite} (\textit{GOES}) soft X-ray monitor on 2017 September 9. The source active region (AR) 12673 was centered at S09W88, very close to the west limb. This active region produced the largest flare in Solar Cycle 24 on September 6 (X9.3) among other large flares in the same period (e.g., M5.5 on September 4, X1.3 on September 7, and X8.2 on September 10). 

The EOVSA images were generated by combining multiple frequency channels over the available 134 frequency channels, to yield 30 spectral windows (spws) over the 3.4--17.9 GHz range. The width of each spw was 160 MHz and the center frequencies were ${f_{GHz}} = 2.92 + n/2$ \citep{2018ApJ...863...83G}. The images were  integrated over 4 seconds to increase the signal-to-noise ratio (SNR), which reduced the temporal cadence to 4 seconds. 

The HXR images were obtained by the image reconstruction software \citep{2002SoPh..210..165S} using the CLEAN algorithm with an integration time of 2 minutes, collimators 6 and 8, which have the nominal FWHM resolutions of 35.3$''$ and 105.8$''$, respectively, and the clean beam width factor of 1.0. The detector choice was made based on the combination of the default detector choice generated by the software at the time of the analysis and the information from the Quicklook per-minute count spectra available from the \textit{RHESSI} Browser \footnote{\url{http://sprg.ssl.berkeley.edu/~tohban/browser/}}.

Figure \ref{fig:development} left panel shows the flare development seen in MW by EOVSA (solid colored contours),
complemented by EUV 131\,\AA{} and 171\,\AA{} (background) images from AIA and the HXR dashed contours from \textit{RHESSI}. The images are plotted for four instances denoted $t_1$--$t_4$ during the time range from 22:44:02 to 22:49:26 UT, which are indicated on the right panel with the black vertical lines in the lighcurves from \textit{GOES}, EOVSA, and \textit{The Gamma-ray Burst Monitor} (\textit{GBM}) on board the \textit{Fermi Gamma-Ray Space Telescope} \citep[\textit{FERMI};][]{2009ApJ...697.1071A} (\href{https://www.dropbox.com/s/i9fi9q1m4gh4xmc/fig1_movie_small.avi?dl=0}{supplemental movie 1} available). In the bottom right panel we show the MW total integrated flux density spectra for the selected four times, which are characterized by an overall increase in flux density without significant spectral changes. We chose this time range  because  the evolving source morphology  was simpler than that during either earlier or later times. 

The flare development throughout the time period shows that the MW high-frequency sources are co-spatial with the HXR nonthermal source (i.e., 25-50 keV) as often observed, indicating the presence of nonthermal electrons in the low solar atmosphere. The centroid of the 25–50 keV HXR source appears to lie very close to the west limb (c.f, Fig. \ref{fig:development}, blue dashed contours), which coincides with the central location of the AR at the photospheric level (S09W88). Although the coarse angular resolution of the detectors we use for imaging does not allow us to derive an accurate height of the HXR source above the limb as in, e.g., \citet{2015ApJ...802...19K}, we assume that it is likely a footpoint source located at chromospheric heights. At progressively lower frequencies, the MW sources extend higher in the corona, indicative of  nonthermal electrons extending to higher heights where the magnetic field is relatively low, which is in line with earlier observations  \citep{2017ApJ...845..135F,2018ApJ...859...17F,2018ApJ...867...81F}. 

Figure \ref{fig:norh_comparison} overlays the EOVSA source contours with the 17\,GHz image (inverted grayscale) obtained with Nobeyama RadioHeliograph \citep[NoRH;][]{1994IEEEP..82..705N}  near the end of the time period, for the same field of view as in Figure\,\ref{fig:development}. 
A similar observation of an extended nonthermal electron population seen in low-frequency MWs was reported by \citet{2018ApJ...863...83G} in the X8.2 flare, which was produced one day later from the same AR. It is interesting to note, however, that the northern footpoint source faintly seen in NoRH image does not appear in the EOVSA images. This is most likely due to the different dynamic ranges of the two instruments.  NoRH has 84 antennas, while EOVSA  has only 13; thus, NoRH has better UV coverage, which results in a better dynamic range (although at lower spatial resolution). As a check, we confirmed that the northern MW source can be faintly seen when a wider, multi-band frequency-synthesis is used for EOVSA imaging, which increases the UV coverage at the expense of frequency resolution. This suggests that the EOVSA images correspond to the more strongly emitting leg of a large, asymmetric loop.  

The HXR low-energy sources (i.e. 6--12 and 12--25 keV) grow over time, very closely following the EUV 131\,\AA\ loops (Figure \ref{fig:development}). We are likely observing a super-hot thermal loop of a few tens of MK, caused either by chromospheric evaporation after the initial particle acceleration seen in the HXR lightcurve, or direct heating from reconnection \citep{2018ApJ...854..178N}. The centroids of the EOVSA MW sources are slightly south of the southern leg of this thermal loop, which is a persistent feature (see \href{https://www.dropbox.com/s/i9fi9q1m4gh4xmc/fig1_movie_small.avi?dl=0}{supplemental movie 1}).
Therefore, the loop with the MW-emitting electrons appears to be  slightly larger and possibly extends above the super-hot thermal loop. This combination of the HXR and MW source morphology  fits the standard flare model: nonthermal electrons are injected from the acceleration region above the super-hot thermal loop, and then travel downward in the magnetic flux tube to emit GS radiation in the corona and bremsstrahlung in the chromosphere.

\section{MW and HXR Analysis of the nonthermal electron distribution}
\label{sec:analysis}

Both HXR and MW emissions are natural outcomes of nonthermal electrons accelerated in flares \citep[][and references therein]{2011SSRv..159..225W}. Observables of these emissions depend on both the nonthermal electron population and local properties of the flaring plasma in regions where those emissions are formed. As a result of the topological diversity of flares, the HXR and MW emissions display a variety of appearances and relationships. In some flares, both emissions are produced by a single population of nonthermal electrons in a single flaring loop \citep[e.g.,][]{2016ApJ...816...62F}.
In other cases, there could be different populations of the nonthermal electrons in distinct flaring loops, thus, different populations can dominate HXR and MW emissions. 
As a result, the nonthermal electron populations forming the two emissions can appear different in terms of spatial and/or energy distributions \citep{1988SoPh..118...49D,1994ApJS...90..599K,2000ApJ...545.1116S}, even though they may have originated from the same acceleration site/process. 

In this flare, the  temporal correlation between HXR and MW lightcurves (seen in Figure \ref{fig:development} right panel) suggests a common origin of accelerated electrons responsible for the two emissions. Spatial relationships are consistent with the standard flare model, as explained in the previous section. Even so, the HXR- and MW-emitting energy ranges of the population may still exhibit dissimilar energy distributions and  evolution thereof. 
Here, we focus on a comparison of the energy distributions of the nonthermal electrons derived from the spatially resolved MW and HXR data. 

\vskip 0.75cm 

\subsection{MW analysis\label{sec:analysis_mw}}

The MW emission in solar flares depends on many  crucial physical parameters including magnetic field strength and  orientation, nonthermal electron energy and angular distribution, and ambient plasma density and temperature. 
To derive those physical parameters, the spatially-resolved spectrum from each pixel in the high-resolution EOVSA images has to be forward-fit with the appropriate cost function. A suitable cost function \citep[][; Nita et al. 2020, in preparation]{2013SoPh..288..549G} employs a numerical fast GS code that accounts for GS and free--free radio emission and absorption \citep{2010ApJ...721.1127F}, since analytical approximations   \citep{1982ApJ...259..350D,1985ARA&A..23..169D} are too limited and too approximate for a meaningful forward fit. This fast GS code is an enhancement of a less accurate numerical Petrosian--Klein (PK) approximation of the exact GS equations \citep{1968Ap&SS...2..171M,1969ApJ...158..753R}, which are highly complicated and computationally slow for our purposes. The fast GS code reduces computation time for GS emission by many orders of magnitude compared to exact calculations, while preserving the needed accuracy.
Performing this model fitting, one can obtain the model fitting parameters over the entire source region at the observational cadence \citep{Fleishman278} in the form of evolving maps of the physical parameters. These parameter maps reveal the spatial distribution and the temporal evolution of the magnetic field and the plasma in the corona.

For the model spectral fitting, we adopt a homogeneous source along the line-of-sight (LOS) and fix the following parameters: plasma temperature, 30 MK; source depth, 5.8 Mm (equivalent to 8 arcsec); an isotropic pitch-angle distribution, and a single power-law electron energy distribution of the form
\begin{equation}
\label{eqn:e_dist}
\begin{aligned}
    \frac{dn(E)}{dE}=A_{0}E^{-\delta} \\
        n =\int\limits_{E_{\min}}^{E_{\max}} \frac{dn(E)}{dE}dE  
\end{aligned}
\end{equation}
where $A_{0}$ is the normalization factor, $\delta$ is the spectral index, $n$ is the number density of the nonthermal electrons with energies between $E_{\min}$ and $E_{\max}$, $E_{\min}$ is the minimum cutoff energy fixed at 17 keV, and $E_{\max}$ is the maximum cutoff energy fixed at 5 MeV.  The initial values of the five free parameters are: nonthermal density, $10^{7}$ cm$^{-3}$; magnetic field strength, 400 G; viewing angle (angle between LOS and the magnetic field line), 60 degrees; thermal density, $10^{10}$ cm$^{-3}$; and $\delta$, 4.5. Although the currently available spectral fitting tool, GSFIT, accepts the $T$ and $E_{\max}$ as free parameters, we found that they are not constrained for this flare; thus, they were fixed as described above.  

The errors of the individual data points, needed to compute the $\chi^2$ metrics, were determined as follows. In each map we selected a region away from the microwave source and computed the rms value of the fluctuations. Then, to take into account the uncertainty introduced by a frequency-dependent spatial resolution of the EOVSA instrument, we added a frequency-dependent systematic uncertainty \citep{2013SoPh..288..549G,Fleishman278}. The actual scatter of the adjacent spectral data points is noticeably smaller than the associated error bars (see examples in Figures \ref{fig:appendix} and \ref{fig:appendix_samplefit}), which implies that the observational errors have been overestimated. For this reason, in what follows we will use a conservative $\chi^2$ upper threshold smaller than conventional values about one. 

Figures \ref{fig:paramovie} and \ref{fig:paratrend} show that the derived physical parameters vary smoothly in both space and time.  
Figure \ref{fig:paramovie} shows a subset of the parameter maps (\href{https://www.dropbox.com/s/a29u09wnea90vh5/fig3_movie.avi?dl=0}{supplemental movie 2}) at four selected times indicated in Figure \ref{fig:development} for: (a) magnetic field strength, (b) thermal electron density, (c) nonthermal electron density, (d) electron energy spectral index, (e) viewing angle, and (f) $\chi^2$ values of the fittings. 
The black contours are the 50\% level of the radio maps at all 30 spws, while the red circles indicate the 50\% contours of the 3.4 GHz images (spw 1) at each time. 
Visual inspection of individual fits suggests  that the spectral fits with $\chi^2 \lesssim 0.1$ are acceptable. Others may not be well fit because of: (1) a complex spectrum inconsistent with the uniform source model and 
contamination of the spectrum due to a sidelobe (see Figure \ref{fig:appendix} (a,b,e)). These ill-fit spectra are excluded from the quantitative analysis. The $\chi^2$ values exceed the threshold in the lower-height part of the sources, which restricts our study to the coronal portion of the flaring loop.

In order to investigate evolution of the parameters, we selected a small area, marked by a red box in the bottom row of Figure \ref{fig:paramovie},  within the 3.4 GHz source that collectively showed $\chi^2$ values less than 0.1 for the longest time. The $\chi^2$ values in the area northward of the red box are lower, but many spectra have fewer or no optically-thick data points, making the fit formally better there but the parameters less reliable than in the red box. 

Figure\,\ref{fig:paratrend} shows evolution of the fit parameters in the red box. The black lines indicate the median values of the parameters from all 25 individual pixels, while the 
gray shade shows the associated error range 
calculated as the standard deviation of the parameters over these 25 pixels. 
In panel (g), we plot the lightcurves of MW 3.4 GHz and HXR 30-100 keV for the reference. The vertical dashed lines correspond to the times $t_1$--$t_4$ shown in Figures \ref{fig:development} and \ref{fig:paramovie}.
During the time range $t_1$--$t_3$, when  $\chi^2 \lesssim 0.1$, we see the following trends:

(1) magnetic field strength is about $\sim$250 G; it does not show significant variations;

(2) thermal electron density varies within $\sim(1-2)\times 10^{10}$ cm$^{-3}$,

(3) nonthermal electron density (above 17 keV) stays relatively constant, at $\sim 10^7$ cm$^{-3}$,

(4) electron energy spectral index hardens significantly, from $14.1 \pm 0.7$ to $5.4 \pm 0.1$,

(5) viewing angle is in the range of 70 to 90 degrees.


To check if these parameter trends are reasonable, we inspect the MW spectral evolution at the center pixel of the red box up to time $t_3$, as shown in Figure \ref{fig:appendix_samplefit}. The flux density of this spatially-resolved spectrum increases by about a factor of 10 in the peak, and more in optically-thin regions as the spectral slope decreases.  This is the expected behavior when the magnetic field strength and nonthermal electron density are both constant, while the spectral index hardens \citep[see Supplemental Movie 2 in][]{Fleishman278}.

The trends (1) and (3) found above supports our initial view of the nature of the MW sources observed during this time period, that they are produced by the electrons accelerated at the acceleration cite and are transported inside the magnetic flux tube. This is in contrast with the result found by \citet{Fleishman278}, where they found the correlated decrease in magnetic field strength and increase in nonthermal electron density in the flare particle acceleration region.

In order to evaluate the effect of fixing a subset of parameters on the spectral fitting results, we perform a separate set of model fitting on the time series of spectra from the central pixel of the red square. 
First, we tested the effect of setting plasma temperature as a free parameter (initial temperature, 5 MK). We found as expected that the fit temperature values are unstable, varying from $\sim1$ MK to $\sim25$ MK, but are smaller than their error values, which means that plasma temperature cannot be well constrained with these data. Even so, we found that the trends in other parameters in Figure\,\ref{fig:paratrend} do not change significantly. We then doubled the source depth to 11.6\,Mm (cf. 5.8\,Mm) and found the derived magnetic field strength drops slightly to $\sim$225 G without significant temporal variation, which is statistically consistent with 250\,G reported above. We then ran the spectral fitting for two alternative values of $E_{\min}$, 10 keV and 30 keV.  For the former, we found that the magnetic field strength decreased with time from $\sim$250~G to $\sim$210~G, while the nonthermal density remained stable at $\sim5\times10^{8}$ cm$^{-3}$. For the latter, the magnetic field remained stable at $\sim$200 G and the nonthermal density did not change from $\sim10^{7}$ cm$^{-3}$. Other fit parameters---thermal electron density, electron energy spectral index, and viewing angle---remained unaffected by changes in those fixed parameters.

One more possible limitation of our model spectral fitting is the assumed isotropic angular distribution of the nonthermal electrons. Although the GS emission certainly depends on the pitch-angle anisotropy \citep{Fl_Meln_2003b}, it is difficult to constrain without imaging spectro-polarimetry data, which is not yet available. Thus, at present we cannot firmly quantify possible bias introduced by the assumption of the isotropic angular distribution.



\subsection{HXR analysis}
\label{sec:analysis_hxr}

We conducted the spectral analysis in HXR using the OSPEX package \citep{2002SoPh..210..165S}. The spectra were obtained from $t_1$--$t_4$ using collimator 3 (which has a good energy resolution and reasonable instrument response matrix), with an integration time of 32 seconds, an energy range 1--106 keV, and with 1/3-keV-wide energy bins. We then fit the spectrum using the thermal (``vth''), thick-target model (``thick2''), and pile-up correction (``pileup\_mod'') functions\footnote{See documentation at: \url{https://hesperia.gsfc.nasa.gov/ssw/packages/spex/idl/object_spex/fit_model_components.txt}}. The equation for the thick-target model is:
\begin{equation}
\label{eqn:thick_target}
    \text{Flux}(\epsilon)=\frac{n_{th}}{4\pi(AU)^2}\frac{1}{mc^2}\int\limits_{\epsilon}^{E_{\max}}\frac{\sigma(\epsilon,E)v}{dE/dt}\int\limits_{\epsilon}^{E_{\max}}F(E_{0})dE_{0}dE
\end{equation}{}
where Flux$(\epsilon)$ is the photon flux at photon energy $\epsilon$, $n_{th}$ is the number density of the thermal plasma, $AU$ is one astronomical unit, $m$ is the electron mass, $c$ is the speed of light, $\sigma(\epsilon,E)$ is the bremsstrahlung cross section from equation (4) of \citet{1997A&A...326..417H}, $v$ is the nonthermal electron speed, and $F(E_{0})$ is the electron flux density distribution function (electrons cm$^{-2}$ s$^{-1}$ keV$^{-1}$), which is returned in the fitting\footnote{See documentation at: \url{https://hesperia.gsfc.nasa.gov/ssw/packages/xray/doc/brm_thick_doc.pdf}}. In order to make this analysis consistent with the MW analysis, we only considered a single power-law and fixed the low energy cutoff of the electron energy distribution to 17 keV. The fitting energy range was $\sim6$ keV to $\sim70$ keV.

In this flare, the instrument had its attenuator state at A0, which made our observation the most affected by the pulse pile-up effect. In order to correct for this effect with some consistency, we have fit each spectrum manually while monitoring that the emission measure from the thermal fit and the first parameter of pile-up correction function (``coefficient to increase or decrease probability of pileup for energies $>$ cutoff'') both increase correspondingly as the low-energy count rates increase during this time period. A sample of the spectral fit results is shown in Figure \ref{fig:appendix_ospex}. The thermal fit returns a plasma temperature of $\sim$32 MK and an emission measure of $\sim2\times10^{46}$cm$^{-3}$, which translates to a density of high $\sim10^{8}$cm$^{-3}$, if we estimate the thermal source volume from Figure \ref{fig:development} by assuming a bicone with a diameter of $\sim50$'' and a height of $\sim50$''$\pi$. This plasma density is about an order of magnitude lower than that obtained from MW spectral fitting. However, we note that the plasma density from MW spectral fitting may not be well constrained, since the spectra at higher heights do not have many optically-thick data points in our frequency range. In fact, it is possible to fit those spectra with lower plasma density while all other parameters are nearly the same as before, if we allow the high-frequency plateau from the free-free emission to be lower than the data points (but still within the error bars). Therefore, it is possible that the plasma density from the MW spectral fitting is overestimated and thus, the actual values could be consistent with the lower, HXR-derived values.  

We see in Figure \ref{fig:development} that the 25--50 keV sources obtained during this time period are most likely footpoint sources, and thus conclude that the fit parameters obtained from the thick-target function can be used to evaluate the powerlaw index of electrons producing the 25--50 keV emission. The time profile of this HXR-derived spectral index is plotted in Figure \ref{fig:paratrend} (d) as a blue line\footnote{Since the spectral index returned by OSPEX fit is that of the electron flux density spectrum, we add 0.5 to the OSPEX values in order to make them comparable to our MW-derived $\delta$, which is that of a number density spectrum. i.e., $n(E)=F(E)/v$, where $n(E)\propto E^{-\delta_{mw}}$, $F(E)\propto E^{-\delta_{HXR}}$, and $v\propto E^{1/2}$.}.

\subsection{Combining the results from MW and HXR analysis}
\label{sec:comparison}

It is apparent from Figure \ref{fig:paratrend} (d) that the $\delta$ values and their evolutions are very different in the corona and the thick-target source. However, we find that our coronal $\delta$ evolution from MW analysis seems to have some correspondence with the light curves of two emissions in Figure \ref{fig:paratrend}(g). The interval up to $t_3$ is divided into two episodes (guided by vertical dashed lines). In interval $t_1$--$t2$, the HXR and MW lightcurves show rapid increase and the coronal $\delta$ also shows rapid hardening from $14.1 \pm 0.7$ to $6.8 \pm 0.3$. In interval $t_2$--$t_3$, the HXR and MW lightcurves show much slower increase (or perhaps none for HXR) and the coronal $\delta$ shows slower hardening as well, from $6.8 \pm 0.3$ to $5.4 \pm 0.1$. In the first episode, the HXR lightcurve's spiky shape suggests particle acceleration and the precipitation of the nonthermal electrons into the chromosphere. At the same time, the significant hardening of our coronal $\delta$ indicates a rapid increase in the number of high-energy nonthermal electrons in the corona, which is reflected by the rapid increase of the coronal MW emission at 3.4 GHz. This correspondence supports our initial view of this flare, where the particle acceleration occurs at or above the 3.4 GHz source, and the accelerated electrons travel downward along the magnetic field lines to emit GS radiation lower in the corona and thick-target bremsstrahung HXR radiation in the lower atmosphere. 

In episode $t_2$--$t_3$ the HXR lightcurve suggests no significant increase in precipitation of nonthermal electrons to the chromosphere compared to the first episode. However, the coronal $\delta$ continues to harden. In order to comprehend this situation, we compare the observed evolution of the coronal electron energy spectrum with the model evolution provided by previous theoretical studies. We use the so-called trap-plus-precipitation (TPP) model \citep{1976MNRAS.176...15M}, which gives the analytical description of the evolution of the energy spectrum of the electrons in the magnetic trap under the influence of electron injection, energy losses due to Coulomb collisions, and precipitation out of the magnetic trap due to the pitch-angle diffusion by Coulomb interactions. This model considers the two extreme cases of (1) initial injection but no continuous injection and (2) no initial injection but continuous injection that is independent of time. 

Let $N(E,t)$ be the total number of electrons per unit energy range in the magnetic trap and $Q(E,t)$ be the number of electron per unit energy injected into the trap in unit time. Assuming that the injection function is in the form of a single power-law with power-law index $\delta$, the initial conditions for case (1) is
\begin{equation}
\label{eqn:pre_initial}
\begin{aligned}
    N(E,0)=KE^{-\delta} \\
    Q(E,t)=0 \text{ for } t>0, \\
\end{aligned}
\end{equation}
and for case (2),
\begin{equation}
\label{eqn:pre_continuous}
\begin{aligned}
    N(E,0)=0 \\
    Q(E,t)=AE^{-\delta}\theta(t) \\
\theta(t)=\begin{cases}
    0 \text{ for } t<0 \\
    1 \text{ for } t>0.  \\
\end{cases}{}
\end{aligned}
\end{equation}
where $A$ and $K$ are constants. The analytical solution of the transport equation for the temporal evolution of $N(E,t)$ given by \citet{1976MNRAS.176...15M} for case (1) is 
\begin{equation}
\begin{aligned}
    N(E,t)=\left(\frac{E_{0}}{E}\right)^{-5/2}N(E_{0},0) \\
    E_{0}=E(1+\tfrac{3}{2}\nu_{0}E^{-3/2}t)^{2/3}. \\
\end{aligned}
\end{equation}
For case (2),
\begin{equation}
\begin{aligned}
    N(E,t)=\frac{AE^{-\delta}}{(\delta +1)\nu_{0}E^{-3/2}}\\
    \{1-(1+\tfrac{3}{2}\nu_{0}E^{-3/2}t)^{-2(\delta +1)/3}\} \\
\end{aligned}
\end{equation}
where $\nu_{0}\approx 5\times10^{-9}n_{th}\,\text{s}^{-1}(\text{keV})^{3/2}$. 

We take the electron energy spectrum observed in the corona at the end of the first episode as the spectrum of ``initial injection'' for case (1) and of ``continuous injection'' for case (2). Therefore, we use $\delta=6.8$ observed at 22:45:38 UT in Eqn. \ref{eqn:pre_initial} and \ref{eqn:pre_continuous}. We also use $n_{th}\sim10^{10}$ cm$^{-3}$ from the observation of thermal electron density shown in Figure \ref{fig:paratrend}. Lastly, we arbitrarily set $A$ and $K$ to be 1, and plot the normalized $N(E,t)$ for two cases over several times during the time period of the second episode (128 seconds). Figure \ref{fig:tpp} (b) shows the result for case (1), initial injection without continuous injection, and panel (c) shows the result for case (2), no initial injection but with continuous injection. Figure \ref{fig:tpp} (a) shows the evolution of the total electron number spectrum derived from microwave data, obtained by multiplying the number density spectrum from Figure \ref{fig:paratrend} (c) and (d) by the total volume occupied by the small red square in Figure \ref{fig:paramovie}.

It is clear from Figure \ref{fig:tpp} that the observed spectral evolution cannot be explained by the case with only initial injection, since the number of electrons in higher energies, up to several hundreds of keV, does not increase in the model. On the other hand, the increase in the number of those high-energy electrons is well captured by the case with continuous injection, although the rate of increase seems to be faster in the model than in the observation. The result of the continuous injection model, as described in \citet{1976MNRAS.176...15M}, is that the spectrum evolves into a double-power law where the spectral index below the break energy $E_{b}$ is harder than that above $E_{b}$ by 1.5, and that the break energy increases with time as $E_{b}=(\frac{3}{2}\nu_{0}t)^{2/3}$. If we calculate the evolution of $\delta$ in the model by assuming a single power-law with $E_{\min}=17$ keV and $E_{\max}$ as the energy up to which the largest change in $\delta$ is observed, which is  $E_{b}=(\frac{3}{2}\nu_{0}t)^{2/3}$ with $t=128$ s ($\sim$450 keV) and is marked by the vertical dashed line in Figure \ref{fig:tpp} (c), then the modeled coronal $\delta$ is 6.5 at $t$=1 s and 5.3 at $t=128$ s. This is in agreement with our observed values of $6.8 \pm 0.3$ at time $t_2$ and $5.4 \pm 0.1$ at time $t_3$. 

This result shows that the observed coronal $\delta$ hardening, which continued into the second episode, is broadly consistent with the TPP model of continuous electron injection into the coronal magnetic trap. However, the fact that there are still some differences, such as the rate of $\delta$ hardening and the values of $\delta$ above several hundred keV between the model and the observation, suggests that our observation cannot be fully explained by this simple model either. The TPP model's assumption of the time-independent power-law injection function is probably too simplistic, and the real injection function is most likely time-dependent and/or more complex than a single power-law. For example, it can be a double power-law, or a single power-law with the high-energy cut-off increasing in time due to a sustained acceleration process. 

Let us discuss further the fact that there is a significant difference between coronal $\delta$ evolution from MW analysis and chromospheric $\delta$ evolution from HXR analysis. Compared to the coronal $\delta$ evolution of $14.1 \pm 0.7$ to $5.4 \pm 0.1$, the chromospheric $\delta$ changes from $5.4 \pm 0.1$ to $5.0 \pm 0.1$. We try to reconcile this observation to our initial picture where the same population of nonthermal electrons, accelerated in the same acceleration episode, is injected into the loop to produce all of the observed MW and HXR emission in this flare. One way to explain the different $\delta$ evolution in HXRs and MWs is if by assuming that the injected electrons have a double power-law energy spectrum and that our observed low-frequency coronal MW emission is more sensitive to the spectrum above a certain break energy $E_{bk}$ ($\neq E_{b}$). This double power-law spectrum could have a low-energy spectral index comparable to the observed HXR-derived $\delta$ up to $E_{bk}$ and a much softer high-energy spectral index (or, a cut-off) above $E_{bk}$. The greater hardening of the MW-derived $\delta$ can then be reconciled by a hardening of the spectrum of the injected electrons only above $E_{bk}$, or by a sustained increase in time of $E_{bk}$ itself. This would be in line with the recent study of \citet{2019ApJ...871...22W}, which conducted the detailed simulation of the GS emission from the electrons with double power-law energy distribution and found that the increased high-energy electrons specified by the second spectral index result in a harder spectral index in the MW flux density spectrum\comma even if the total number of electrons does not change significantly.

\section{The evolution of the total energy of the nonthermal electrons in the corona}
\label{sec:evolution}

In the previous section, we introduced the general picture of the evolution of the energy spectrum of the nonthermal electrons injected into the flaring loop in this flare. We did so by interpreting the temporal behavior of the parameters in a small representative volume in the context of the existing theory of electron transport in the corona. We now explore the collective behavior of the nonthermal electrons evolving in the flaring loop. Specifically, we calculate the total energy of nonthermal electrons contained in the corona, defined by the 50\% contour of 3.4 GHz image (red contour in Figure \ref{fig:paramovie}), and plot this energy against time, to obtain the evolution of the total nonthermal electron energy contained in the corona. To do so, for each time frame we proceed with the following steps:

(1) exclude all ill-fitted pixels within the 50\% contour of the 3.4 GHz source,

(2) calculate the nonthermal electron energy density in all well-fit pixels,

(3) calculate the weighted mean of (2), then

(4) multiply this weighted mean of the energy density by the total volume within the 50\% contour of the 3.4 GHz source assuming a depth 5.8 Mm.

In step (1), we identify a pixel as ill-fitted if (1) the magnetic field solution is hitting its predefined upper limit of 3,000 G, (2) the number density solution is hitting its predefined lower limit of $10^{3}$~cm$^{-3}$, (3) the spectral index solution is hitting its predefined lower limit of 4 or upper limit of 15, or (4) $\chi^2 > 0.1$. Figure \ref{fig:e_prof5} (c) shows the percentage of the well-fit pixels selected for the analysis with respect to the total number of pixels within 50\% contour of 3.4 GHz source. 

In step (2), we calculate the total energy contained in the coronal MW 3.4 GHz source, which we proposed in the previous section to be sensitive to the electrons that have energies higher than a certain $E_{bk}$. Therefore, we intentionally ``cut'' the observed energy distribution at $E_{bk}$. and obtain the energy density only \emph{above} $E_{bk}$ by using
\begin{equation}
\begin{aligned}
\label{eqn:e_density}
    \varepsilon =  1.6\times10^{-9}\Big(\frac{\delta-1}{\delta-2}\Big)n_{E>E_{min}}E_{min} \\
    \boldsymbol{{E_{bk}}=\alpha E_{min}} \\
    n_{E>E_{bk}}=\alpha^{1-\delta}n_{E>E_{\min}} \\
    \varepsilon_{E>E_{bk}} = 1.6\times10^{-9}\Big(\frac{\delta-1}{\delta-2}\Big)n_{E>E_{bk}}E_{bk} \\
    = \boldsymbol{1.6\times10^{-9}\Big(\frac{\delta-1}{\delta-2}\Big)n_{E>E_{min}}E_{min}\alpha^{2-\delta}}
\end{aligned}
\end{equation}
where $n_{E>E_{min}}$ is $n$, the number density we obtained in Section \ref{sec:analysis_mw} (same as Eqn. \ref{eqn:e_dist}), $E_{min}$ is 17 keV, which was used in obtaining $n$, and $\alpha$ is the factor by which a certain $E_{bk}$ is larger than 17 keV. We do not know the exact value of $E_{bk}$, but we adopt 70 keV for this analysis, since this is the upper limit of the fitting energy range for HXR spectral analysis that shows generally unchanged spectral indices over time.

In step (3), the weighted mean of the energy density is calculated by
\begin{equation}
    <\varepsilon>_{weighted}=\frac{ \sum\limits_{i=1}^{n}  w_{\varepsilon,i}\varepsilon_{i}}{\sum\limits_{i=1}^{n} w_{\varepsilon,i}}
\end{equation}
where $w_{\varepsilon,i}$ is the weight of energy density for $i^{th}$ well-fit pixel, calculated by $1/\varepsilon_{err,i}^{2}$ where $\varepsilon_{err,i}$ is the error in energy density.

Finally, in step (4) we calculate the total volume within the 50\% contour of the 3.4 GHz source by multiplying the total number of pixels within that contour by the 4 arcsec$^{2}$ pixel area and the source depth of 5.8 Mm, as adopted in section \ref{sec:analysis_mw}. The evolution of this volume is plotted in Figure \ref{fig:e_prof5} (b).

Figure \ref{fig:e_prof5}(a) shows the evolution of the total instantaneous energy of the nonthermal electrons having energies $>70$ keV contained within the 50\% contour of the observed MW 3.4 GHz sources. Figure~\ref{fig:e_prof5}\,(d) shows, for reference, the lightcurves from Figure \ref{fig:paratrend} (g), and the four vertical dashed lines mark the same time boundaries as in Figures \ref{fig:development}, \ref{fig:paramovie} and \ref{fig:paratrend}. Figure \ref{fig:e_prof5}(a) shows that, overall, there is a significant increase in the total instantaneous energy of electrons $>70$~keV contained in the coronal source. The energies corresponding to each time boundary are $3.8 \pm 0.5 \times 10^{17}$ erg at $t_1$, $4.9 \pm 0.8 \times 10^{21}$ erg at $t_2$, $3.9 \pm 0.2 \times 10^{24}$ erg at $t_3$, and $7.6 \pm 0.5 \times 10^{24}$ erg at $t_4$. We note that the fraction of well-fit pixels within the source significantly decreases after $t_3$, so the results after this time may not be as reliable as in the preceding time intervals. However, up to $t_3$, at least half of the pixels within the 50\% contour of 3.4 GHz sources are considered for the analysis, so our results can be taken with more confidence during this time. From Figure \ref{fig:e_prof5}(a) it is observed that the total energy reaches about $10^{22}$~erg by time $t_2$, and increases by $\sim$3 orders of magnitude during episode $t_2$--$t_3$.

Although this analysis has been done assuming that the observed radio emission during our time of interest is produced by nonthermal electrons, it is possible that, at least during the time when the inferred spectral index is very soft (e.g., during episode 1), the emission is due to thermal electrons. In fact, most of the time during episode 1, we find that the observed spectra within the red box in Figure \ref{fig:paratrend} could be reproduced without the need for a nonthermal electron distribution. Instead, they could be produced by gyrosynchrotron emission generated by electrons with a $\sim$20 MK thermal distribution in a source with the same magnetic field, thermal electron density, and viewing angle. Therefore, the result in Figure \ref{fig:e_prof5} (a) during episode $t_1$--$t_2$ should be taken as the extreme case of considering all MW-emitting electrons to be nonthermal. However, purely thermal emission is excluded for $t_2$--$t_3$, since many pixels start showing spectra that are too hard in their optically-thin sides to be explained by thermal gyrosynchrotron emission. Therefore, we believe that our results during $t_2$--$t_3$ truly reflect the rise in the energy of nonthermal electrons with energies $>$70 keV in the corona. 

\section{Summary \& Discussion}
\label{sec:summary}

In this study, we analyzed the evolution of the nonthermal electrons accelerated during the impulsive phase of the M1.2 flare on 2017 September 9. We focused on a $\sim$6-minute period when a significant increase of MW emission was observed compared to the level of increase in HXR emission. We used multi-wavelength observations to evaluate the overall spatial distribution of electrons in the flare, and combined it with the total energy and energy distribution of electrons derived from the combination of MW and HXR analysis. In particular, our MW analysis was conducted using the new technique of numerical forward-fitting of spatially-resolved MW spectra derived from multi-frequency images from EOVSA. This enabled the quantitative calculation of the spatially-resolved evolution of the nonthermal electrons in the corona. The summary of the main results from this work is the following.

(1) The comparison of EUV-loops, the locations of low- and high-energy HXR emission sources, and the distribution of MW images from 3.4-18 GHz suggest that the spatial distribution of the nonthermal electrons in this flare generally fits the traditional standard flare model morphology. We infer that the electrons are accelerated in a region located above the hot loop visible in AIA 131\,\AA\ and RHESSI 6-12~keV images and are injected into a somewhat larger ``nonthermal" flare loop connecting the low-frequency coronal MW sources with the co-spatial 20-50~keV HXR and 17~GHz MW sources in the lower atmosphere.

(2) The comparison of the spectral index of the nonthermal electrons derived from HXR and MW analysis reveals, however, that their values and evolution are significantly different. More specifically, the spectral index of the nonthermal electrons associated with the coronal 3.4 GHz MW source undergoes much faster hardening than that associated with the footpoint HXR source. The former hardens from $14.1 \pm 0.7$ to $5.4 \pm 0.1$ and the latter changes from $5.4 \pm 0.1$ to $5.0 \pm 0.1$ over the same period of 128 seconds.  Because the energy range of electrons producing HXR and MW emission differ, we interpret this discrepancy as reflecting a different spectral evolution in different parts of the energy spectrum.

(3) Our findings vividly show that the high-energy tail of the nonthermal electron distribution, which is responsible for the MW emission, underwent more significant evolution compared with the low-energy counterpart of the distribution for this flare.
In-depth analysis focusing on the spectral evolution of the coronal electron population suggests that there was a sustained acceleration and continued injection of nonthermal electrons in the corona even when there was no significant signature for that in the HXR lightcurve for a period of time. The difference in the spectral evolution is reconciled by adopting that the energy spectrum of this injected population has a double power-law or a break-down spectrum above an evolving (rising in time) break energy $E_{bk}$ in the form of a cut-off. The population with energies higher than the break energy, to which the MW emission is more sensitive than the HXR emission, have undergone greater spectral hardening, perhaps, due to the sustained acceleration.

(4) Based on this picture of the evolution of the energy spectrum of the nonthermal electrons injected into the flaring loop, we estimated the evolution of the total instantaneous nonthermal ($>70$~keV) electron energy contained in a coronal volume enclosed by the coronal 3.4 GHz MW source. We find a significant increase of several orders of magnitude in the total energy of these electrons contained in the coronal source during the $\sim$4 minute period of interest of the flare impulsive phase.

An interesting observation, shown in Figure \ref{fig:e_prof5}(a), is that the total instantaneous nonthermal electron energy contained in the coronal source continues to increase during the period we studied. Under the assumption of a single loop, it is interesting to compare this evolution with the development of the total energy flux deposited by HXR-emitting nonthermal electrons into the lower atmosphere. We plot the evolution of the total energy flux of HXR-emitting electrons in Figure \ref{fig:hxren}, using the result of the analysis from Section \ref{sec:analysis_hxr}. We use the modification of  Eqn. \ref{eqn:e_density}, $1.6\times10^{-9}\Big(\frac{\delta-1}{\delta-2}\Big)\mathcal{F}E_{min}$, where $\mathcal{F}$ is the total electron flux obtained from the thick-target spectral fit. Although this cannot be directly compared with Figure \ref{fig:e_prof5}(a) because of the different units, this clearly indicates that the evolution of the total energy of nonthermal electrons in the lower atmosphere is different from that in the corona. 

A consistency check of the single-population scenario can be performed by assuming a single power-law distribution that extends from tens of keV to hundreds of keV, covering both HXR-emitting and MW-emitting population of energetic electrons, at the time when the spectral indices from two analysis match, around $t_3$ (see Figure \ref{fig:paratrend} (d)). In this analysis we use a simple relation $\mathcal{F}(E)=n(E)\overline{v}A$ where $\mathcal{F}(E)$ is the nonthermal electron flux distribution from HXR analysis, $n(E)$ is the nonthermal electron density distribution from MW analysis, $\overline{v}$ is the mean speed of nonthermal electrons, and $A$ is the area of the thick-target HXR emission. We obtain from OSPEX analysis that $\mathcal{F}\sim10^{35}$ electrons/s, estimate $A$ from the 50\% contour of 18 GHz MW image in Figure \ref{fig:development} left panels (circular area with $\sim$10$''$ diameter), and $\overline{v}=\frac{\int_{E_{min}}^{E_{max}}vn(E)dE}{\int_{E_{min}}^{E_{max}}n(E)dE}\sim0.3\,c$ from the energy distribution. This yields $n\sim3\times10^{7}$~cm$^{-3}$, which roughly agrees with the value of $n\sim10^{7}$~cm$^{-3}$ we obtained from MW analysis at this time.

An alternative explanation for the difference in $\delta$ development is that we are observing two electron populations belonging to different loop systems. For instance, our HXR-producing population may be reflecting the evolution of the nonthermal electrons in a small loop unresolved by RHESSI. This will allow the $\delta$ evolution in the corona and lower atmosphere to be unrelated, as in our observation. In fact, \citet{2018ApJ...852...32K} revealed the presence of an extended ''HXR-invisible'' nonthermal electron population outside of the traditional flare geometry at one time during another flare by using the \textit{RHESSI} and the EOVSA data, and this snap-shot model required two separate loops to simultaneously reproduce the observed low-frequency MW emission and HXR/high-frequency MW emission. If we adopt a two-loop scenario for this study, however, the two loops must be dynamically connected since time profiles of low-frequency MWs at higher heights are very well correlated with higher-frequency MWs co-spatial with the 25-50~keV HXR source. 

Lastly, we note that electron pitch-angle anisotropy may change our result in Figure \ref{fig:paratrend} and possibly affect our conclusions. Differences in the inverted nonthermal electron energy distribution for isotropic vs. anisotropic (e.g., beam-like) distributions have been reported in MW both observationally \citep{2008ApJ...677.1367A,2002ApJ...580L.185M,2000ApJ...543..457L} and in simulations \citep{2010ApJ...721.1127F}. For HXR, the bremsstrahlung cross section is also pitch-angle dependent (Equation (2) BN in \citealt{1959RvMP...31..920K}). Therefore, an anisotropic electron distribution would produce a HXR photon spectrum that deviates from our results which assume an isotropic electron distribution \citep[see, e.g., discussions in][]{2004ApJ...613.1233M,2012ApJ...750...35C}. However, exploring the effects of different pitch-angle distributions is beyond the scope of this study. 

Although the origin of the striking dissimilarities in the evolution of the spectral indices of the nonthermal electrons in the corona and the chromosphere are open to debate, it is important to note that these differences are only revealed through the spatially-resolved analysis of the evolving coronal MW sources below $\sim$10 GHz. The continued electron injection and hardening revealed by the coronal MW analysis seems to affect only the highest electron energies, and therefore lacks the expected counterpart signature in the HXR source. 
Since the location of the MW peak frequency is most sensitive to the magnetic field strength of the source, this characteristic informs us about nonthermal electrons higher in the corona (weaker magnetic field), where the HXR analysis becomes increasingly difficult due to the scarcity of the target plasma for bremsstrahlung. 

\section*{Conflict of Interest Statement}

The authors declare that the research was conducted in the absence of any commercial or financial relationships that could be construed as a potential conflict of interest.

\section*{Author Contributions}

NK processed the raw data, performed the fitting, initial analysis of the results, and wrote the initial draft. GF developed the fitting methodology. DG led the construction and commissioning of the EOVSA, developed
the strategy for taking and calibrating the data for microwave spectroscopy. GN developed the GSFIT package used in the automated fitting. BC and SY developed the microwave spectral imaging and self-calibration strategy. All authors involved in the discussion and contributed to manuscript preparation.

\section*{Funding}
EOVSA operation is supported by NSF grant AST-1910354. Natsuha Kuroda was partially supported by the NASA Living With a Star Jack Eddy Postdoctoral Fellowship Program, administered by UCAR’s Cooperative Programs
for the Advancement of Earth System Science (CPAESS) under award \#NNX16AK22G.  G.F, D.G., G.N., B.C., and S.Y. are supported by NSF grants AGS-1654382, AST-1910354, AGS-1817277, AGS-1743321, and NASA grants 80NSSC18K0667, NNX17AB82G, 80NSSC18K1128, and 80NSSC19K0068 to New Jersey Institute of Technology.

\section*{Acknowledgments}
The authors thank the scientists and engineers who helped design and build EOVSA, especially Gordon Hurford, Stephen White, James McTiernan, Wes Grammer, Kjell Nelin, and Tim Bastian.

\section*{Supplemental Data}
The movie corresponding to Figure \ref{fig:development} is available as a supplemental material fig1\_movie.avi. The movie corresponding to Figure \ref{fig:paramovie} is available as a supplemental material fig3\_movie.avi.

\section*{Data Availability Statement}
EOVSA data are freely available from
\url{http://ovsa.njit.edu/data-browsing.html}.
Fully processed EOVSA spectral imaging data in IDL save format can be downloaded from \url{https://www.dropbox.com/s/6dk40smvea8f66f/eovsa_20170909T223518-231518_r2_XX.sav?dl=0}. The package for microwave data fitting, GSFIT, is included in the publicly available library
SolarSoftWare under the Packages
category at \url{http://www.lmsal.com/solarsoft/ssw_packages_info.html}.

\newpage

\appendix
\renewcommand{\thefigure}{A\arabic{figure}}
\section*{Samples of spatially-resolved spectra and their fits}
\label{sec:appendix}

We show in here the sample of spatially-resolved MW spectra (black asterisks) and the fit results (green lines). The examples of successful fits are (c) and (d) and those of unsuccessful fits are (a), (b), and (e). The pixel selected for (d) and (e) is one of the pixels in the small red box shown in Figure~\ref{fig:paramovie}. (a) through (d) show the results from 22:46:14~UT, a time between $t_2$ and $t_3$, and (e) shows the result from the same pixel as (d) but at $t_4$, at the end of our fitting time period. 

\begin{figure}[ht!]
\includegraphics[scale=.6]{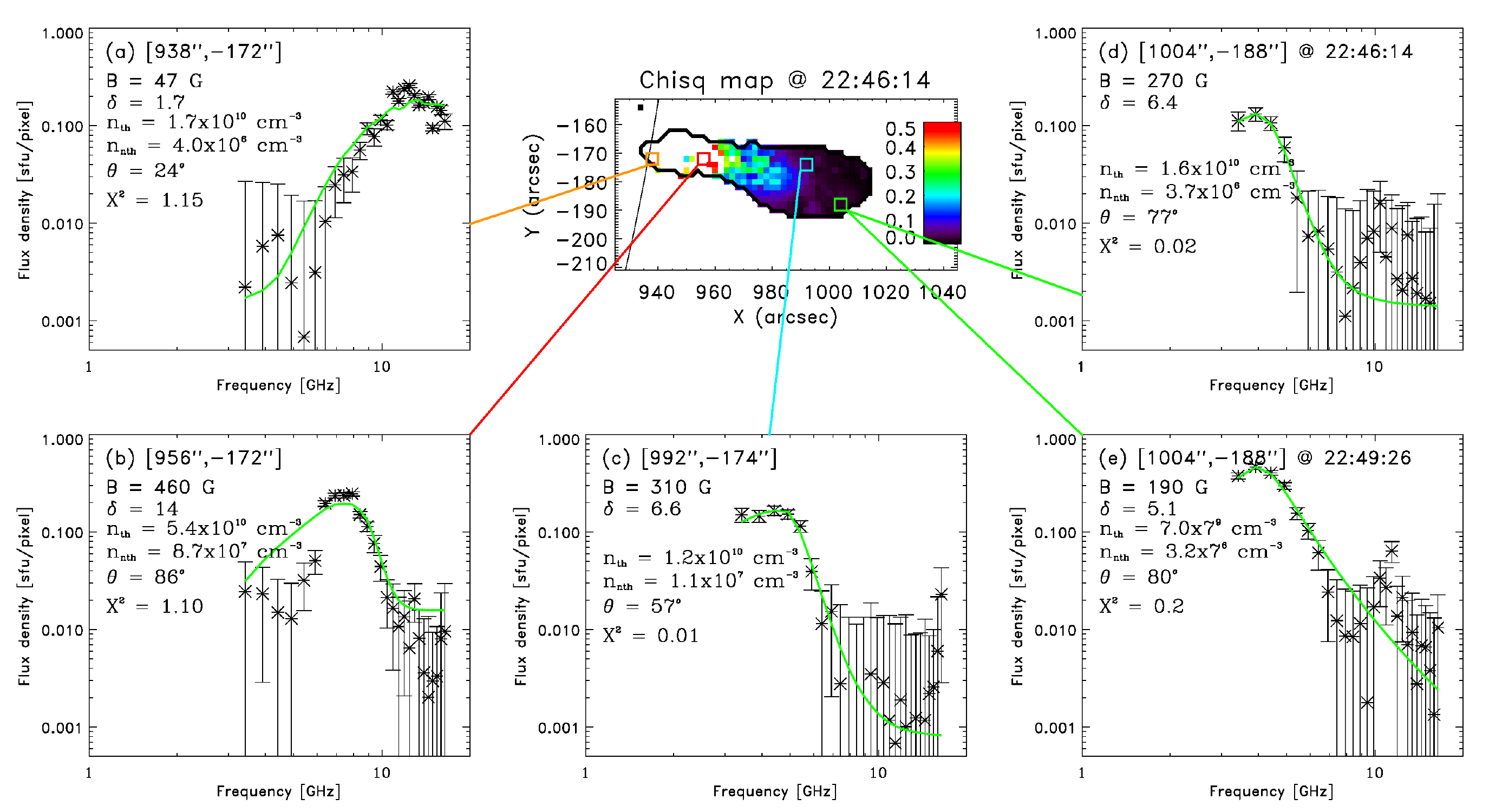}
\caption{(a) An unsuccessful fit due to the high peak frequency leading to the shortage of good data points in the optically-thin frequency range. Note the unrealistically low magnetic field value of 47~G, far lower than fits of surrounding pixels. (b) An unsuccessful fit due to the unusually narrow peak shape of the observed spectrum. (c,d) Examples of successful fits. Note that the magnetic field strength reasonably decreases with height. The $\chi^{2}$ value is $<0.1$, which is the criterion used in the quantitative analysis in Section 3. (e) An unsuccessful fit due to the contamination of the spectrum by a sidelobe of the high-frequency source at $\sim$ 11 GHz. The result is shown from the same pixel as (d) but at later time, at 22:49:26. This type of unsuccessful fit was more prelavent toward the end of our fitting time period as the flare emission intensity increased. Note that with this fit, the spectral index may be reported harder than the actual value. The $\chi^{2}$ value is again $>0.1$.}\label{fig:appendix}
\end{figure}

\newpage

\bibliographystyle{frontiersinSCNS_ENG_HUMS} 
\bibliography{20170909flare}

\begin{thebibliography}{48}
\providecommand{\natexlab}[1]{#1}
\expandafter\ifx\csname urlstyle\endcsname\relax
  \providecommand{\doi}[1]{doi:\discretionary{}{}{}#1}\else
  \providecommand{\doi}{doi:\discretionary{}{}{}\begingroup
  \urlstyle{rm}\Url}\fi
\providecommand{\selectlanguage}[1]{\relax}
\providecommand{\bibAnnoteFile}[1]{%
  \IfFileExists{#1}{\begin{quotation}\noindent\textsc{Key:} #1\\
  \textsc{Annotation:}\ \input{#1}\end{quotation}}{}}
\providecommand{\bibAnnote}[2]{%
  \begin{quotation}\noindent\textsc{Key:} #1\\
  \textsc{Annotation:}\ #2\end{quotation}}

\bibitem[{{Altyntsev} et~al.(2008){Altyntsev}, {Fleishman}, {Huang}, and
  {Melnikov}}]{2008ApJ...677.1367A}
{Altyntsev}, A.~T., {Fleishman}, G.~D., {Huang}, G.~L., and {Melnikov}, V.~F.
  (2008).
\newblock {A Broadband Microwave Burst Produced by Electron Beams}.
\newblock \emph{\apj} 677, 1367--1377.
\newblock \doi{10.1086/528841}
\bibAnnoteFile{2008ApJ...677.1367A}

\bibitem[{{Aschwanden} et~al.(2016){Aschwanden}, {Holman}, {O'Flannagain},
  {Caspi}, {McTiernan}, and {Kontar}}]{2016ApJ...832...27A}
{Aschwanden}, M.~J., {Holman}, G., {O'Flannagain}, A., {Caspi}, A.,
  {McTiernan}, J.~M., and {Kontar}, E.~P. (2016).
\newblock {Global Energetics of Solar Flares. III. Nonthermal Energies}.
\newblock \emph{\apj} 832, 27.
\newblock \doi{10.3847/0004-637X/832/1/27}
\bibAnnoteFile{2016ApJ...832...27A}

\bibitem[{{Atwood} et~al.(2009){Atwood}, {Abdo}, {Ackermann}, {Althouse},
  {Anderson}, {Axelsson} et~al.}]{2009ApJ...697.1071A}
{Atwood}, W.~B., {Abdo}, A.~A., {Ackermann}, M., {Althouse}, W., {Anderson},
  B., {Axelsson}, M., et~al. (2009).
\newblock {The Large Area Telescope on the Fermi Gamma-Ray Space Telescope
  Mission}.
\newblock \emph{\apj} 697, 1071--1102.
\newblock \doi{10.1088/0004-637X/697/2/1071}
\bibAnnoteFile{2009ApJ...697.1071A}

\bibitem[{{Chen} and {Bastian}(2012)}]{2012ApJ...750...35C}
{Chen}, B. and {Bastian}, T.~S. (2012).
\newblock {The Role of Inverse Compton Scattering in Solar Coronal Hard X-Ray
  and {\ensuremath{\gamma}}-Ray Sources}.
\newblock \emph{\apj} 750, 35.
\newblock \doi{10.1088/0004-637X/750/1/35}
\bibAnnoteFile{2012ApJ...750...35C}

\bibitem[{{Dennis}(1988)}]{1988SoPh..118...49D}
{Dennis}, B.~R. (1988).
\newblock {Solar Flare Hard X-Ray Observations}.
\newblock \emph{\solphys} 118, 49--94.
\newblock \doi{10.1007/BF00148588}
\bibAnnoteFile{1988SoPh..118...49D}

\bibitem[{{Dulk}(1985)}]{1985ARA&A..23..169D}
{Dulk}, G.~A. (1985).
\newblock {Radio emission from the sun and stars.}
\newblock \emph{\araa} 23, 169--224.
\newblock \doi{10.1146/annurev.aa.23.090185.001125}
\bibAnnoteFile{1985ARA&A..23..169D}

\bibitem[{{Dulk} and {Marsh}(1982)}]{1982ApJ...259..350D}
{Dulk}, G.~A. and {Marsh}, K.~A. (1982).
\newblock {Simplified expressions for the gyrosynchrotron radiation from mildly
  relativistic, nonthermal and thermal electrons}.
\newblock \emph{\apj} 259, 350--358.
\newblock \doi{10.1086/160171}
\bibAnnoteFile{1982ApJ...259..350D}

\bibitem[{{Emslie} et~al.(2012){Emslie}, {Dennis}, {Shih}, {Chamberlin},
  {Mewaldt}, {Moore} et~al.}]{2012ApJ...759...71E}
{Emslie}, A.~G., {Dennis}, B.~R., {Shih}, A.~Y., {Chamberlin}, P.~C.,
  {Mewaldt}, R.~A., {Moore}, C.~S., et~al. (2012).
\newblock {Global Energetics of Thirty-eight Large Solar Eruptive Events}.
\newblock \emph{\apj} 759, 71.
\newblock \doi{10.1088/0004-637X/759/1/71}
\bibAnnoteFile{2012ApJ...759...71E}

\bibitem[{Fleishman et~al.(2020)Fleishman, Gary, Chen, Kuroda, Yu, and
  Nita}]{Fleishman278}
Fleishman, G.~D., Gary, D.~E., Chen, B., Kuroda, N., Yu, S., and Nita, G.~M.
  (2020).
\newblock Decay of the coronal magnetic field can release sufficient energy to
  power a solar flare.
\newblock \emph{Science} 367, 278--280.
\newblock \doi{10.1126/science.aax6874}
\bibAnnoteFile{Fleishman278}

\bibitem[{{Fleishman} et~al.(2011){Fleishman}, {Kontar}, {Nita}, and
  {Gary}}]{Fl_etal_2011}
{Fleishman}, G.~D., {Kontar}, E.~P., {Nita}, G.~M., and {Gary}, D.~E. (2011).
\newblock {A Cold, Tenuous Solar Flare: Acceleration Without Heating}.
\newblock \emph{\apjl} 731, L19.
\newblock \doi{10.1088/2041-8205/731/1/L19}
\bibAnnoteFile{Fl_etal_2011}

\bibitem[{{Fleishman} et~al.(2013){Fleishman}, {Kontar}, {Nita}, and
  {Gary}}]{2013ApJ...768..190F}
{Fleishman}, G.~D., {Kontar}, E.~P., {Nita}, G.~M., and {Gary}, D.~E. (2013).
\newblock {Probing Dynamics of Electron Acceleration with Radio and X-Ray
  Spectroscopy, Imaging, and Timing in the 2002 April 11 Solar Flare}.
\newblock \emph{\apj} 768, 190.
\newblock \doi{10.1088/0004-637X/768/2/190}
\bibAnnoteFile{2013ApJ...768..190F}

\bibitem[{{Fleishman} and {Kuznetsov}(2010)}]{2010ApJ...721.1127F}
{Fleishman}, G.~D. and {Kuznetsov}, A.~A. (2010).
\newblock {Fast Gyrosynchrotron Codes}.
\newblock \emph{\apj} 721, 1127--1141.
\newblock \doi{10.1088/0004-637X/721/2/1127}
\bibAnnoteFile{2010ApJ...721.1127F}

\bibitem[{{Fleishman} et~al.(2018{\natexlab{a}}){Fleishman}, {Loukitcheva},
  {Kopnina}, {Nita}, and {Gary}}]{2018ApJ...867...81F}
{Fleishman}, G.~D., {Loukitcheva}, M.~A., {Kopnina}, V.~Y., {Nita}, G.~M., and
  {Gary}, D.~E. (2018{\natexlab{a}}).
\newblock {The Coronal Volume of Energetic Particles in Solar Flares as
  Revealed by Microwave Imaging}.
\newblock \emph{\apj} 867, 81.
\newblock \doi{10.3847/1538-4357/aae0f6}
\bibAnnoteFile{2018ApJ...867...81F}

\bibitem[{{Fleishman} and {Melnikov}(2003)}]{Fl_Meln_2003b}
{Fleishman}, G.~D. and {Melnikov}, V.~F. (2003).
\newblock {Gyrosynchrotron Emission from Anisotropic Electron Distributions}.
\newblock \emph{\apj} 587, 823--835.
\newblock \doi{10.1086/368252}
\bibAnnoteFile{Fl_Meln_2003b}

\bibitem[{{Fleishman} et~al.(2009){Fleishman}, {Nita}, and
  {Gary}}]{2009ApJ...698L.183F}
{Fleishman}, G.~D., {Nita}, G.~M., and {Gary}, D.~E. (2009).
\newblock {Dynamic Magnetography of Solar Flaring Loops}.
\newblock \emph{\apjl} 698, L183--L187.
\newblock \doi{10.1088/0004-637X/698/2/L183}
\bibAnnoteFile{2009ApJ...698L.183F}

\bibitem[{{Fleishman} et~al.(2017){Fleishman}, {Nita}, and
  {Gary}}]{2017ApJ...845..135F}
{Fleishman}, G.~D., {Nita}, G.~M., and {Gary}, D.~E. (2017).
\newblock {A Large-scale Plume in an X-class Solar Flare}.
\newblock \emph{\apj} 845, 135.
\newblock \doi{10.3847/1538-4357/aa81d4}
\bibAnnoteFile{2017ApJ...845..135F}

\bibitem[{{Fleishman} et~al.(2016{\natexlab{a}}){Fleishman}, {Nita}, {Kontar},
  and {Gary}}]{2016ApJ...826...38F}
{Fleishman}, G.~D., {Nita}, G.~M., {Kontar}, E.~P., and {Gary}, D.~E.
  (2016{\natexlab{a}}).
\newblock {Narrowband Gyrosynchrotron Bursts: Probing Electron Acceleration in
  Solar Flares}.
\newblock \emph{\apj} 826, 38.
\newblock \doi{10.3847/0004-637X/826/1/38}
\bibAnnoteFile{2016ApJ...826...38F}

\bibitem[{{Fleishman} et~al.(2018{\natexlab{b}}){Fleishman}, {Nita}, {Kuroda},
  {Jia}, {Tong}, {Wen} et~al.}]{2018ApJ...859...17F}
{Fleishman}, G.~D., {Nita}, G.~M., {Kuroda}, N., {Jia}, S., {Tong}, K., {Wen},
  R.~R., et~al. (2018{\natexlab{b}}).
\newblock {Revealing the Evolution of Non-thermal Electrons in Solar Flares
  Using 3D Modeling}.
\newblock \emph{\apj} 859, 17.
\newblock \doi{10.3847/1538-4357/aabae9}
\bibAnnoteFile{2018ApJ...859...17F}

\bibitem[{{Fleishman} et~al.(2016{\natexlab{b}}){Fleishman}, {Xu}, {Nita}, and
  {Gary}}]{2016ApJ...816...62F}
{Fleishman}, G.~D., {Xu}, Y., {Nita}, G.~N., and {Gary}, D.~E.
  (2016{\natexlab{b}}).
\newblock {Validation of the Coronal Thick Target Source Model}.
\newblock \emph{\apj} 816, 62.
\newblock \doi{10.3847/0004-637X/816/2/62}
\bibAnnoteFile{2016ApJ...816...62F}

\bibitem[{{Gary} et~al.(2018){Gary}, {Chen}, {Dennis}, {Fleishman}, {Hurford},
  {Krucker} et~al.}]{2018ApJ...863...83G}
{Gary}, D.~E., {Chen}, B., {Dennis}, B.~R., {Fleishman}, G.~D., {Hurford},
  G.~J., {Krucker}, S., et~al. (2018).
\newblock {Microwave and Hard X-Ray Observations of the 2017 September 10 Solar
  Limb Flare}.
\newblock \emph{\apj} 863, 83.
\newblock \doi{10.3847/1538-4357/aad0ef}
\bibAnnoteFile{2018ApJ...863...83G}

\bibitem[{{Gary} et~al.(2013){Gary}, {Fleishman}, and
  {Nita}}]{2013SoPh..288..549G}
{Gary}, D.~E., {Fleishman}, G.~D., and {Nita}, G.~M. (2013).
\newblock {Magnetography of Solar Flaring Loops with Microwave Imaging
  Spectropolarimetry}.
\newblock \emph{\solphys} 288, 549--565.
\newblock \doi{10.1007/s11207-013-0299-3}
\bibAnnoteFile{2013SoPh..288..549G}

\bibitem[{{Glesener} and {Fleishman}(2018)}]{2018ApJ...867...84G}
{Glesener}, L. and {Fleishman}, G.~D. (2018).
\newblock {Electron Acceleration and Jet-facilitated Escape in an M-class Solar
  Flare on 2002 August 19}.
\newblock \emph{\apj} 867, 84.
\newblock \doi{10.3847/1538-4357/aacefe}
\bibAnnoteFile{2018ApJ...867...84G}

\bibitem[{{Haug}(1997)}]{1997A&A...326..417H}
{Haug}, E. (1997).
\newblock {On the use of nonrelativistic bremsstrahlung cross sections in
  astrophysics.}
\newblock \emph{\aap} 326, 417--418
\bibAnnoteFile{1997A&A...326..417H}

\bibitem[{{Hoyng} et~al.(1981){Hoyng}, {Duijveman}, {Machado}, {Rust},
  {Svestka}, {Boelee} et~al.}]{1981ApJ...246L.155H}
{Hoyng}, P., {Duijveman}, A., {Machado}, M.~E., {Rust}, D.~M., {Svestka}, Z.,
  {Boelee}, A., et~al. (1981).
\newblock {Origin and Location of the Hard X-Ray Emission in a Two-Ribbon
  Flare}.
\newblock \emph{\apjl} 246, L155.
\newblock \doi{10.1086/183574}
\bibAnnoteFile{1981ApJ...246L.155H}

\bibitem[{{Koch} and {Motz}(1959)}]{1959RvMP...31..920K}
{Koch}, H.~W. and {Motz}, J.~W. (1959).
\newblock {Bremsstrahlung Cross-Section Formulas and Related Data}.
\newblock \emph{Reviews of Modern Physics} 31, 920--955.
\newblock \doi{10.1103/RevModPhys.31.920}
\bibAnnoteFile{1959RvMP...31..920K}

\bibitem[{{Krucker} and {Battaglia}(2014)}]{2014ApJ...780..107K}
{Krucker}, S. and {Battaglia}, M. (2014).
\newblock {Particle Densities within the Acceleration Region of a Solar Flare}.
\newblock \emph{\apj} 780, 107.
\newblock \doi{10.1088/0004-637X/780/1/107}
\bibAnnoteFile{2014ApJ...780..107K}

\bibitem[{{Krucker} et~al.(2010){Krucker}, {Hudson}, {Glesener}, {White},
  {Masuda}, {Wuelser} et~al.}]{2010ApJ...714.1108K}
{Krucker}, S., {Hudson}, H.~S., {Glesener}, L., {White}, S.~M., {Masuda}, S.,
  {Wuelser}, J.~P., et~al. (2010).
\newblock {Measurements of the Coronal Acceleration Region of a Solar Flare}.
\newblock \emph{\apj} 714, 1108--1119.
\newblock \doi{10.1088/0004-637X/714/2/1108}
\bibAnnoteFile{2010ApJ...714.1108K}

\bibitem[{{Krucker} et~al.(2015){Krucker}, {Saint-Hilaire}, {Hudson},
  {Haberreiter}, {Martinez-Oliveros}, {Fivian} et~al.}]{2015ApJ...802...19K}
{Krucker}, S., {Saint-Hilaire}, P., {Hudson}, H.~S., {Haberreiter}, M.,
  {Martinez-Oliveros}, J.~C., {Fivian}, M.~D., et~al. (2015).
\newblock {Co-Spatial White Light and Hard X-Ray Flare Footpoints Seen Above
  the Solar Limb}.
\newblock \emph{\apj} 802, 19.
\newblock \doi{10.1088/0004-637X/802/1/19}
\bibAnnoteFile{2015ApJ...802...19K}

\bibitem[{{Kundu} et~al.(1994){Kundu}, {White}, {Gopalswamy}, and
  {Lim}}]{1994ApJS...90..599K}
{Kundu}, M.~R., {White}, S.~M., {Gopalswamy}, N., and {Lim}, J. (1994).
\newblock {Millimeter, Microwave, Hard X-Ray, and Soft X-Ray Observations of
  Energetic Electron Populations in Solar Flares}.
\newblock \emph{\apjs} 90, 599.
\newblock \doi{10.1086/191881}
\bibAnnoteFile{1994ApJS...90..599K}

\bibitem[{{Kuroda} et~al.(2018){Kuroda}, {Gary}, {Wang}, {Fleishman}, {Nita},
  and {Jing}}]{2018ApJ...852...32K}
{Kuroda}, N., {Gary}, D.~E., {Wang}, H., {Fleishman}, G.~D., {Nita}, G.~M., and
  {Jing}, J. (2018).
\newblock {Three-dimensional Forward-fit Modeling of the Hard X-Ray and
  Microwave Emissions of the 2015 June 22 M6.5 Flare}.
\newblock \emph{\apj} 852, 32.
\newblock \doi{10.3847/1538-4357/aa9d98}
\bibAnnoteFile{2018ApJ...852...32K}

\bibitem[{{Lee} and {Gary}(2000)}]{2000ApJ...543..457L}
{Lee}, J. and {Gary}, D.~E. (2000).
\newblock {Solar Microwave Bursts and Injection Pitch-Angle Distribution of
  Flare Electrons}.
\newblock \emph{\apj} 543, 457--471.
\newblock \doi{10.1086/317080}
\bibAnnoteFile{2000ApJ...543..457L}

\bibitem[{{Lemen} et~al.(2012){Lemen}, {Title}, {Akin}, {Boerner}, {Chou},
  {Drake} et~al.}]{2012SoPh..275...17L}
{Lemen}, J.~R., {Title}, A.~M., {Akin}, D.~J., {Boerner}, P.~F., {Chou}, C.,
  {Drake}, J.~F., et~al. (2012).
\newblock {The Atmospheric Imaging Assembly (AIA) on the Solar Dynamics
  Observatory (SDO)}.
\newblock \emph{\solphys} 275, 17--40.
\newblock \doi{10.1007/s11207-011-9776-8}
\bibAnnoteFile{2012SoPh..275...17L}

\bibitem[{{Lin} et~al.(2002){Lin}, {Dennis}, {Hurford}, {Smith}, {Zehnder},
  {Harvey} et~al.}]{2002SoPh..210....3L}
{Lin}, R.~P., {Dennis}, B.~R., {Hurford}, G.~J., {Smith}, D.~M., {Zehnder}, A.,
  {Harvey}, P.~R., et~al. (2002).
\newblock {The Reuven Ramaty High-Energy Solar Spectroscopic Imager (RHESSI)}.
\newblock \emph{\solphys} 210, 3--32.
\newblock \doi{10.1023/A:1022428818870}
\bibAnnoteFile{2002SoPh..210....3L}

\bibitem[{{Lin} and {Hudson}(1971)}]{1971SoPh...17..412L}
{Lin}, R.~P. and {Hudson}, H.~S. (1971).
\newblock {10 100 keV electron acceleration and emission from solar flares}.
\newblock \emph{\solphys} 17, 412--435.
\newblock \doi{10.1007/BF00150045}
\bibAnnoteFile{1971SoPh...17..412L}

\bibitem[{{Massone} et~al.(2004){Massone}, {Emslie}, {Kontar}, {Piana},
  {Prato}, and {Brown}}]{2004ApJ...613.1233M}
{Massone}, A.~M., {Emslie}, A.~G., {Kontar}, E.~P., {Piana}, M., {Prato}, M.,
  and {Brown}, J.~C. (2004).
\newblock {Anisotropic Bremsstrahlung Emission and the Form of Regularized
  Electron Flux Spectra in Solar Flares}.
\newblock \emph{\apj} 613, 1233--1240.
\newblock \doi{10.1086/423127}
\bibAnnoteFile{2004ApJ...613.1233M}

\bibitem[{{Masuda} et~al.(1994){Masuda}, {Kosugi}, {Hara}, {Tsuneta}, and
  {Ogawara}}]{1994Natur.371..495M}
{Masuda}, S., {Kosugi}, T., {Hara}, H., {Tsuneta}, S., and {Ogawara}, Y.
  (1994).
\newblock {A loop-top hard X-ray source in a compact solar flare as evidence
  for magnetic reconnection}.
\newblock \emph{\nat} 371, 495--497.
\newblock \doi{10.1038/371495a0}
\bibAnnoteFile{1994Natur.371..495M}

\bibitem[{{Melnikov} et~al.(2002){Melnikov}, {Shibasaki}, and
  {Reznikova}}]{2002ApJ...580L.185M}
{Melnikov}, V.~F., {Shibasaki}, K., and {Reznikova}, V.~E. (2002).
\newblock {Loop-Top Nonthermal Microwave Source in Extended Solar Flaring
  Loops}.
\newblock \emph{\apjl} 580, L185--L188.
\newblock \doi{10.1086/345587}
\bibAnnoteFile{2002ApJ...580L.185M}

\bibitem[{{Melrose}(1968)}]{1968Ap&SS...2..171M}
{Melrose}, D.~B. (1968).
\newblock {The Emission and Absorption of Waves by Charged Particles in
  Magnetized Plasmas}.
\newblock \emph{\apss} 2, 171--235.
\newblock \doi{10.1007/BF00651567}
\bibAnnoteFile{1968Ap&SS...2..171M}

\bibitem[{{Melrose} and {Brown}(1976)}]{1976MNRAS.176...15M}
{Melrose}, D.~B. and {Brown}, J.~C. (1976).
\newblock {Precipitation in trap models for solar hard X-ray bursts.}
\newblock \emph{\mnras} 176, 15--30.
\newblock \doi{10.1093/mnras/176.1.15}
\bibAnnoteFile{1976MNRAS.176...15M}

\bibitem[{{Nakajima} et~al.(1994){Nakajima}, {Nishio}, {Enome}, {Shibasaki},
  {Takano}, {Hanaoka} et~al.}]{1994IEEEP..82..705N}
{Nakajima}, H., {Nishio}, M., {Enome}, S., {Shibasaki}, K., {Takano}, T.,
  {Hanaoka}, Y., et~al. (1994).
\newblock {The Nobeyama radioheliograph.}
\newblock \emph{IEEE Proceedings} 82, 705--713
\bibAnnoteFile{1994IEEEP..82..705N}

\bibitem[{{Ning} et~al.(2018){Ning}, {Chen}, {Wu}, {Su}, {Tian}, {Li}
  et~al.}]{2018ApJ...854..178N}
{Ning}, H., {Chen}, Y., {Wu}, Z., {Su}, Y., {Tian}, H., {Li}, G., et~al.
  (2018).
\newblock {Two-stage Energy Release Process of a Confined Flare with Double HXR
  Peaks}.
\newblock \emph{\apj} 854, 178.
\newblock \doi{10.3847/1538-4357/aaaa69}
\bibAnnoteFile{2018ApJ...854..178N}

\bibitem[{{Nita} et~al.(2016){Nita}, {Hickish}, {MacMahon}, and
  {Gary}}]{2016JAI.....541009N}
{Nita}, G.~M., {Hickish}, J., {MacMahon}, D., and {Gary}, D.~E. (2016).
\newblock {EOVSA Implementation of a Spectral Kurtosis Correlator for Transient
  Detection and Classification}.
\newblock \emph{Journal of Astronomical Instrumentation} 5, 1641009-7366.
\newblock \doi{10.1142/S2251171716410099}
\bibAnnoteFile{2016JAI.....541009N}

\bibitem[{{Nitta} and {Kosugi}(1986)}]{1986SoPh..105...73N}
{Nitta}, N. and {Kosugi}, T. (1986).
\newblock {Energy of Microwave-Emitting Electrons and Hard X-Ray / Microwave
  Source Model in Solar Flares}.
\newblock \emph{\solphys} 105, 73--85.
\newblock \doi{10.1007/BF00156378}
\bibAnnoteFile{1986SoPh..105...73N}

\bibitem[{{Ramaty}(1969)}]{1969ApJ...158..753R}
{Ramaty}, R. (1969).
\newblock {Gyrosynchrotron Emission and Absorption in a Magnetoactive Plasma}.
\newblock \emph{\apj} 158, 753.
\newblock \doi{10.1086/150235}
\bibAnnoteFile{1969ApJ...158..753R}

\bibitem[{{Schwartz} et~al.(2002){Schwartz}, {Csillaghy}, {Tolbert}, {Hurford},
  {McTiernan}, and {Zarro}}]{2002SoPh..210..165S}
{Schwartz}, R.~A., {Csillaghy}, A., {Tolbert}, A.~K., {Hurford}, G.~J.,
  {McTiernan}, J., and {Zarro}, D. (2002).
\newblock {RHESSI Data Analysis Software: Rationale and Methods}.
\newblock \emph{\solphys} 210, 165--191.
\newblock \doi{10.1023/A:1022444531435}
\bibAnnoteFile{2002SoPh..210..165S}

\bibitem[{{Silva} et~al.(2000){Silva}, {Wang}, and
  {Gary}}]{2000ApJ...545.1116S}
{Silva}, A. V.~R., {Wang}, H., and {Gary}, D.~E. (2000).
\newblock {Correlation of Microwave and Hard X-Ray Spectral Parameters}.
\newblock \emph{\apj} 545, 1116--1123.
\newblock \doi{10.1086/317822}
\bibAnnoteFile{2000ApJ...545.1116S}

\bibitem[{{White} et~al.(2011){White}, {Benz}, {Christe}, {F{\'a}rn{\'\i}k},
  {Kundu}, {Mann} et~al.}]{2011SSRv..159..225W}
{White}, S.~M., {Benz}, A.~O., {Christe}, S., {F{\'a}rn{\'\i}k}, F., {Kundu},
  M.~R., {Mann}, G., et~al. (2011).
\newblock {The Relationship Between Solar Radio and Hard X-ray Emission}.
\newblock \emph{\ssr} 159, 225--261.
\newblock \doi{10.1007/s11214-010-9708-1}
\bibAnnoteFile{2011SSRv..159..225W}

\bibitem[{{Wu} et~al.(2019){Wu}, {Chen}, {Ning}, {Kong}, and
  {Lee}}]{2019ApJ...871...22W}
{Wu}, Z., {Chen}, Y., {Ning}, H., {Kong}, X., and {Lee}, J. (2019).
\newblock {Gyrosynchrotron Emission Generated by Nonthermal Electrons with the
  Energy Spectra of a Broken Power Law}.
\newblock \emph{\apj} 871, 22.
\newblock \doi{10.3847/1538-4357/aaf474}
\bibAnnoteFile{2019ApJ...871...22W}

\end{thebibliography}



\renewcommand{\thefigure}{\arabic{figure}}
\setcounter{figure}{0}
\begin{figure*}[ht!]
\includegraphics[scale=.5]{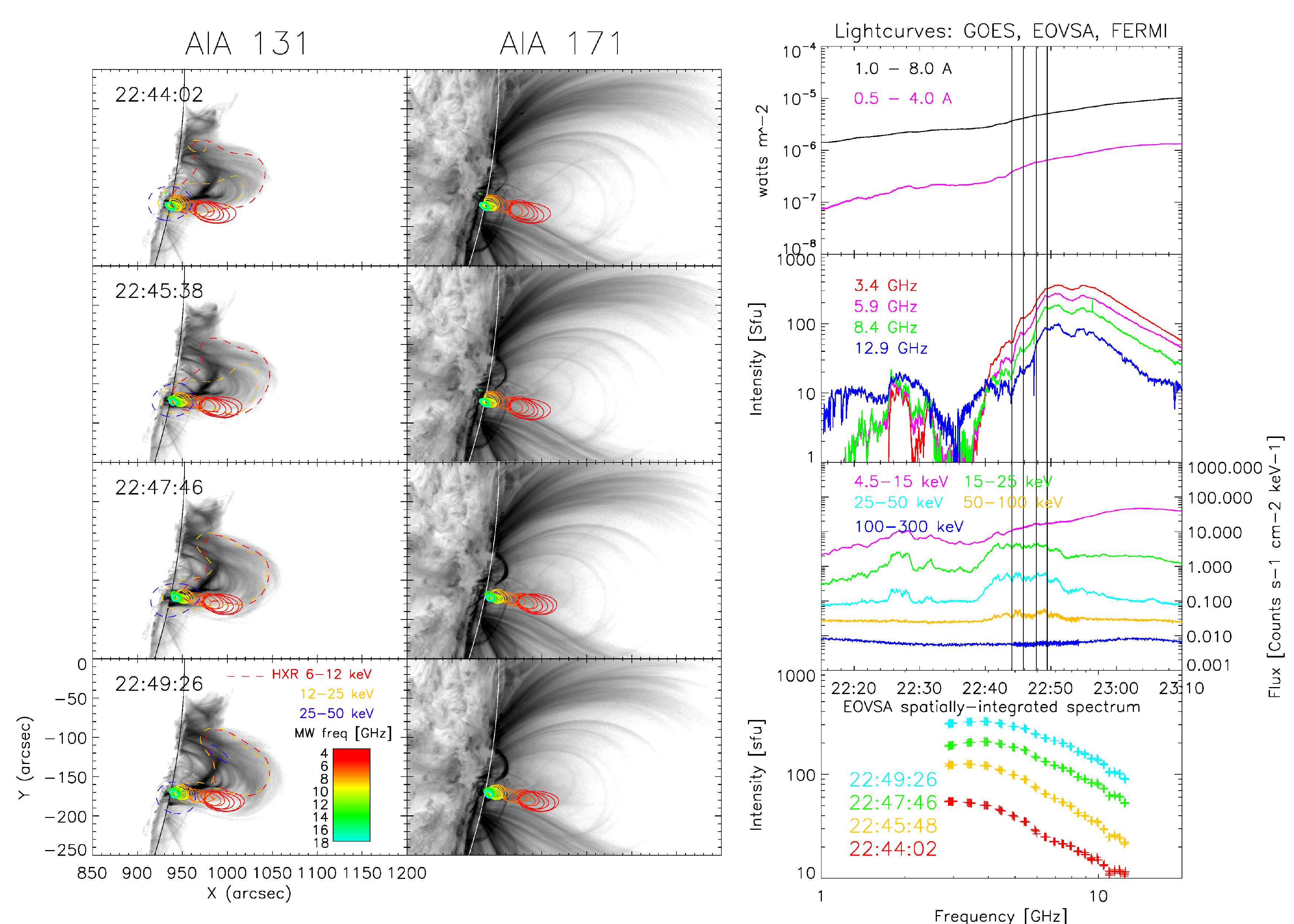}
\caption{Left panel: The flare development seen in MWs by EOVSA (solid contours [The color contours show 50\% of the maximum intensity of each of 30 spw images]) and HXR from \textit{RHESSI} (dashed contours), plotted on EUV 131\,\AA\ and 171\,\AA\ (background inverted grayscale) from \textit{SDO/AIA}, at times $t_1$--$t_4$ (22:44:02--22:49:26 UT), indicated by four black vertical lines on the right panel. Right panel, from the top: The soft X-ray, MW, and HXR lightcurves from \textit{GOES}, EOVSA, and \textit{FERMI}, respectively, and the total flux density spectra from EOVSA at four times images on the left panel (movie available). All contours are at 50$\%$ of the maximum intensity of each image.} \label{fig:development}
\end{figure*}

\begin{figure}[ht!]
\includegraphics[scale=1.2]{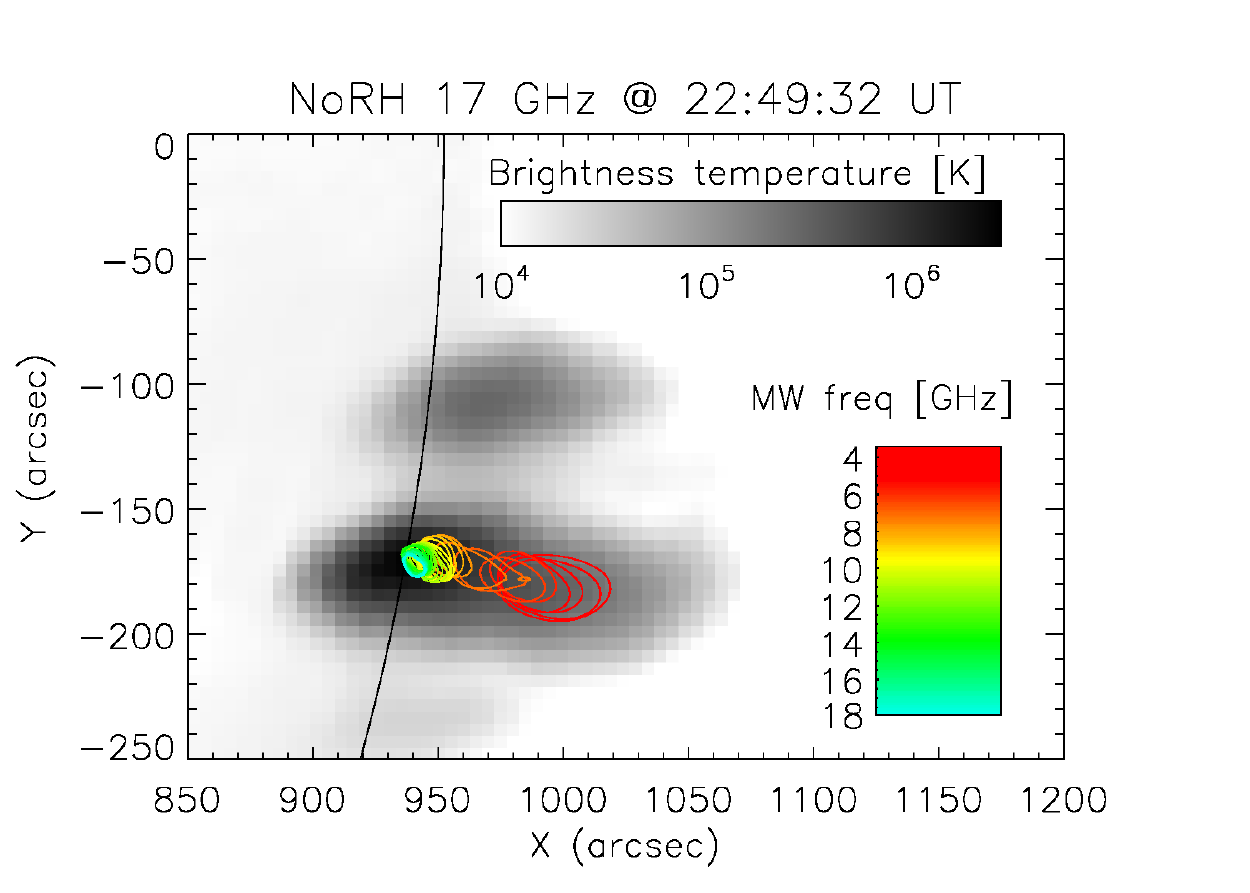}
\caption{The EOVSA sources overlaid on the NoRH 17 GHz image (inverted grayscale) taken near the end of our time of interest, 22:49:26. The field of view is the same as in Figure \ref{fig:development}. EOVSA's low frequency coverage reveals the large spatial extent of the nonthermal electron population in the corona with respect to height. At the same time, the higher dynamic range of the NoRH image at a single frequency reveals a possible asymmetry in the magnetic field strength of the loop containing these electrons. }\label{fig:norh_comparison}
\end{figure}

\begin{figure*}[ht!]
\includegraphics[scale=.3]{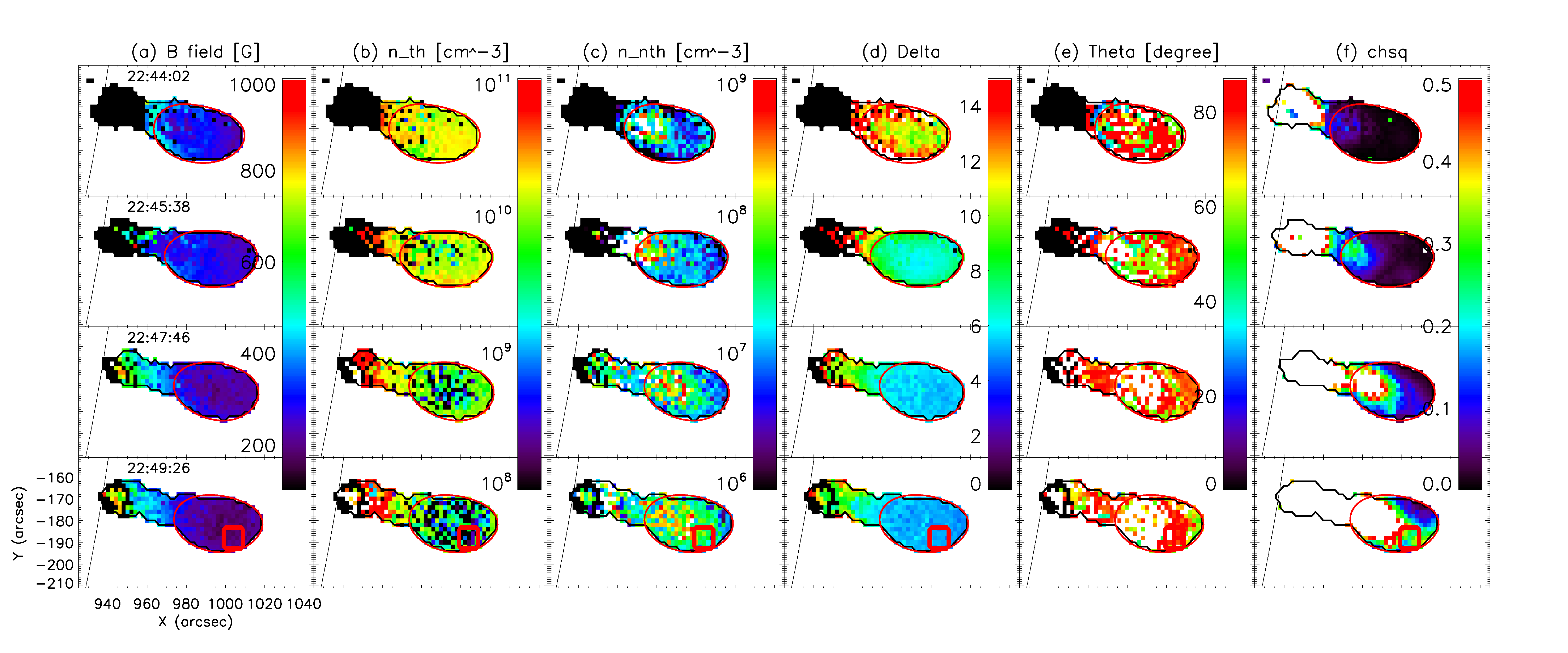}
\caption{Snapshots of the coronal magnetic field and plasma parameter movie created by forward-fitting the spatially-resolved spectra of EOVSA images at every 4 seconds (movie available). The four times are the same times as shown in Figure \ref{fig:development}. The black outline marks the union of the 50\% contours of the original images at all 30 spws at each time, and the red circles indicate the 50\% contours of the lowest frequency images (spw 1, 3.4~GHz) at each time. The time profiles of the parameters in the small red box are shown in Figure \ref{fig:paratrend}. 
}\label{fig:paramovie}
\end{figure*}

\begin{figure}[ht!]
\includegraphics[scale=.7]{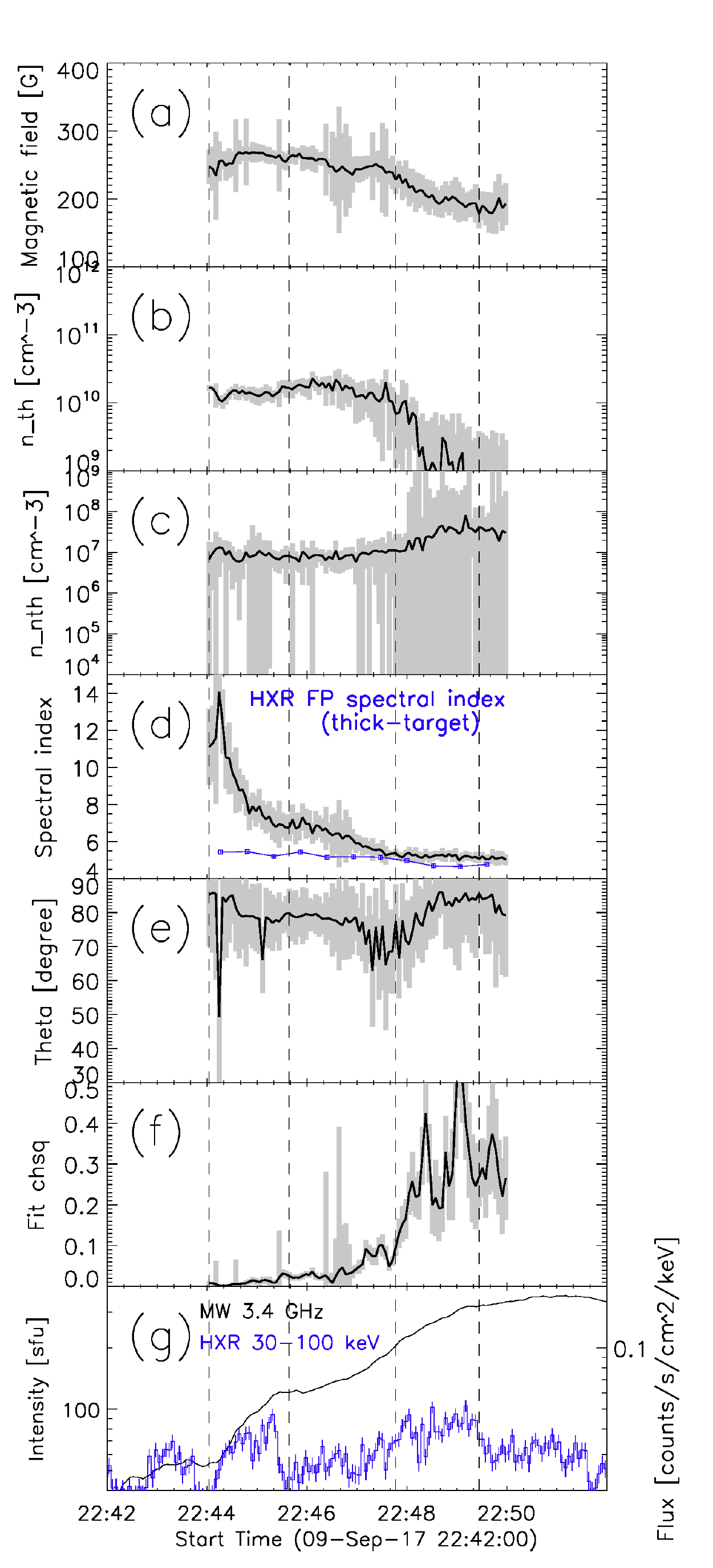}
\caption{(a-f) Time profiles of the parameters from 25 pixels within the small red square marked in Figure \ref{fig:paramovie}: gray shade indicates the error bar of each parameter at each instance calculated as the standard deviation of the parameters over 25 pixels, black for their median. The spectral index inferred from the HXR analysis, interpreted as the index of electrons emitting in the lower atmosphere in Section \ref{sec:analysis_hxr}, is plotted in blue in (d). (g) The lowest-frequency MW lightcurve and the high-energy HXR lightcurve from \textit{RHESSI} for reference, plotted on the same log scale. The dashed lines indicate time $t_1$--$t_4$.}\label{fig:paratrend}
\end{figure}

\begin{figure}[ht!]
\includegraphics{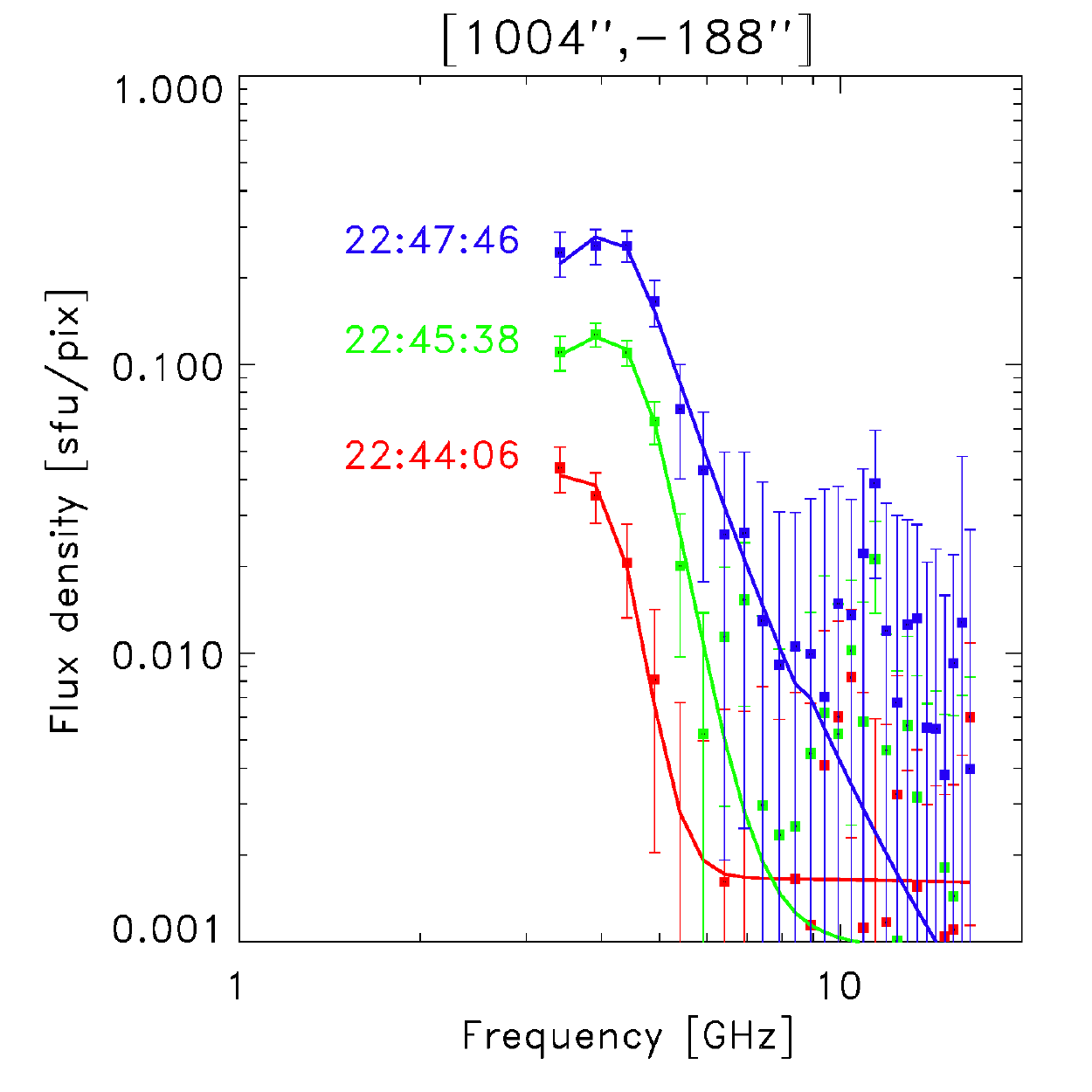}
\caption{An example of the temporal evolution of the spectra and their fits, taken at one pixel inside the red box in Figure \ref{fig:paramovie}. The three times correspond to times $t_1$--$t_3$.}\label{fig:appendix_samplefit}
\end{figure}

\begin{figure}[ht!]
\includegraphics[angle=90,scale=.15]{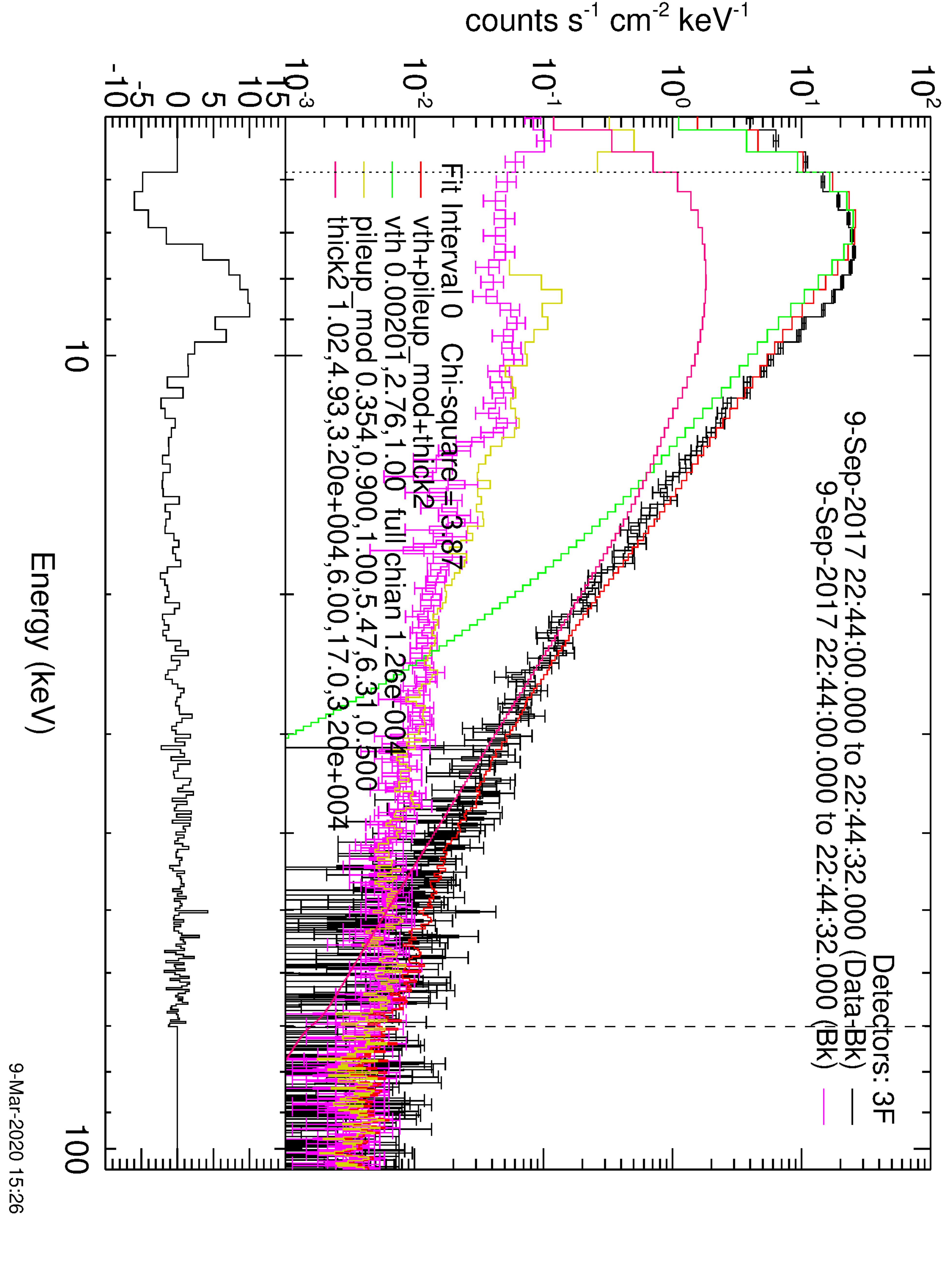}
\caption{A example of the HXR spectral fit from OSPEX, at 22:44:16 UT.}\label{fig:appendix_ospex}
\end{figure}

\begin{figure}[ht!]
\includegraphics[scale=.7]{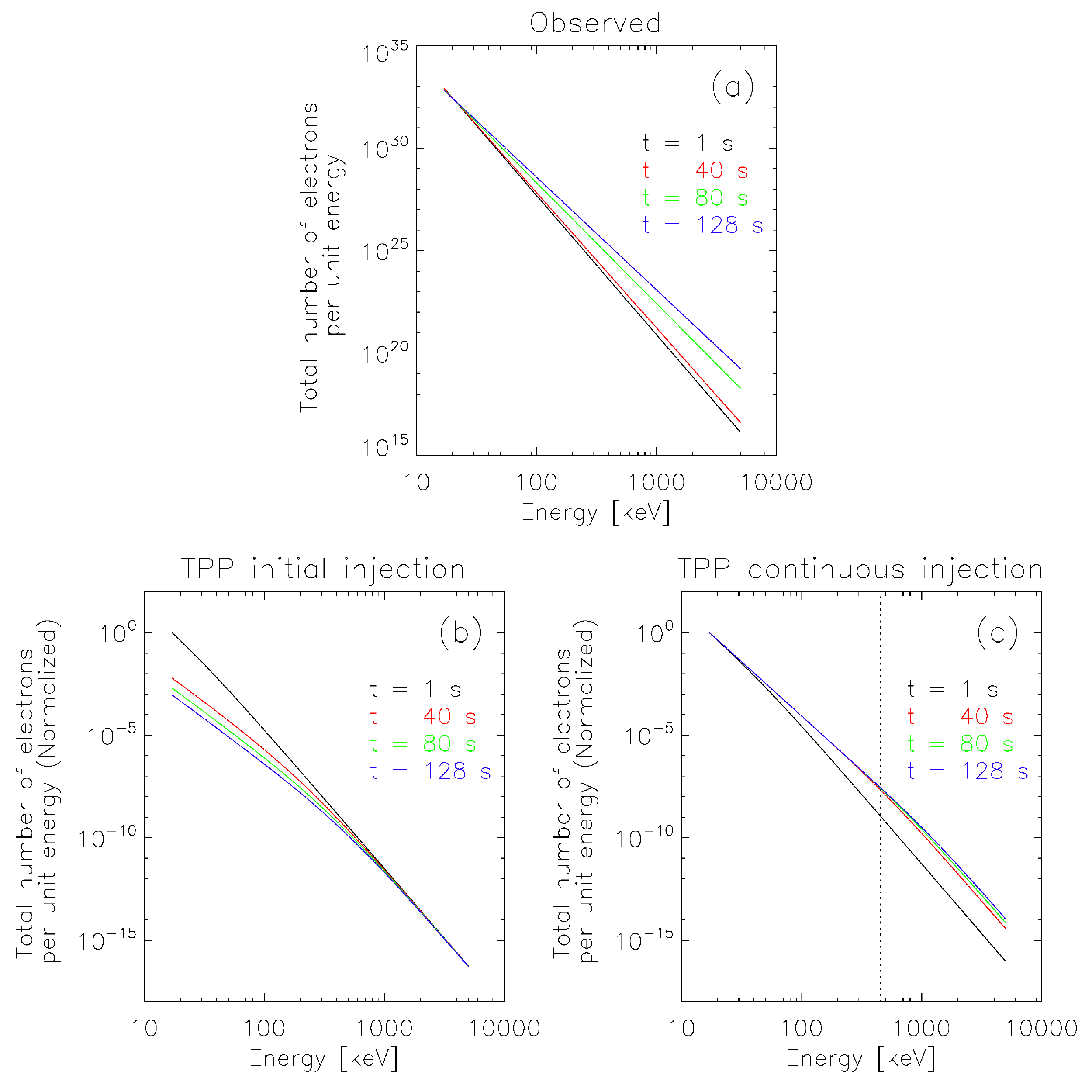}
\caption{(a) The evolution of total electron number spectrum deduced from the MW fits, obtained by multiplying the number density spectrum from Figure \ref{fig:paratrend} (c) and (d) by the total volume occupied by the red box in Figure \ref{fig:paramovie}. (b,c) The modeled evolution of the total electron number spectrum (normalized) for the trap-plus-precipitation (TPP) model by \citet{1976MNRAS.176...15M}, over several times during the period of 22:45:38 to 22:47:46 UT, when the coronal $\delta$ from MW analysis shows a continued hardening despite a lack of apparent change in electron injection deduced from the HXR lightcurve. (b) Initial injection without continuous injection. (c) No initial injection but with continuous injection. The vertical dotted line marks the analytical value of the energy up to which the largest change in the low-energy spectral index is observed (see Section \ref{sec:comparison}).
}\label{fig:tpp}
\end{figure}

\begin{figure}[ht!]
\includegraphics[scale=1.3]{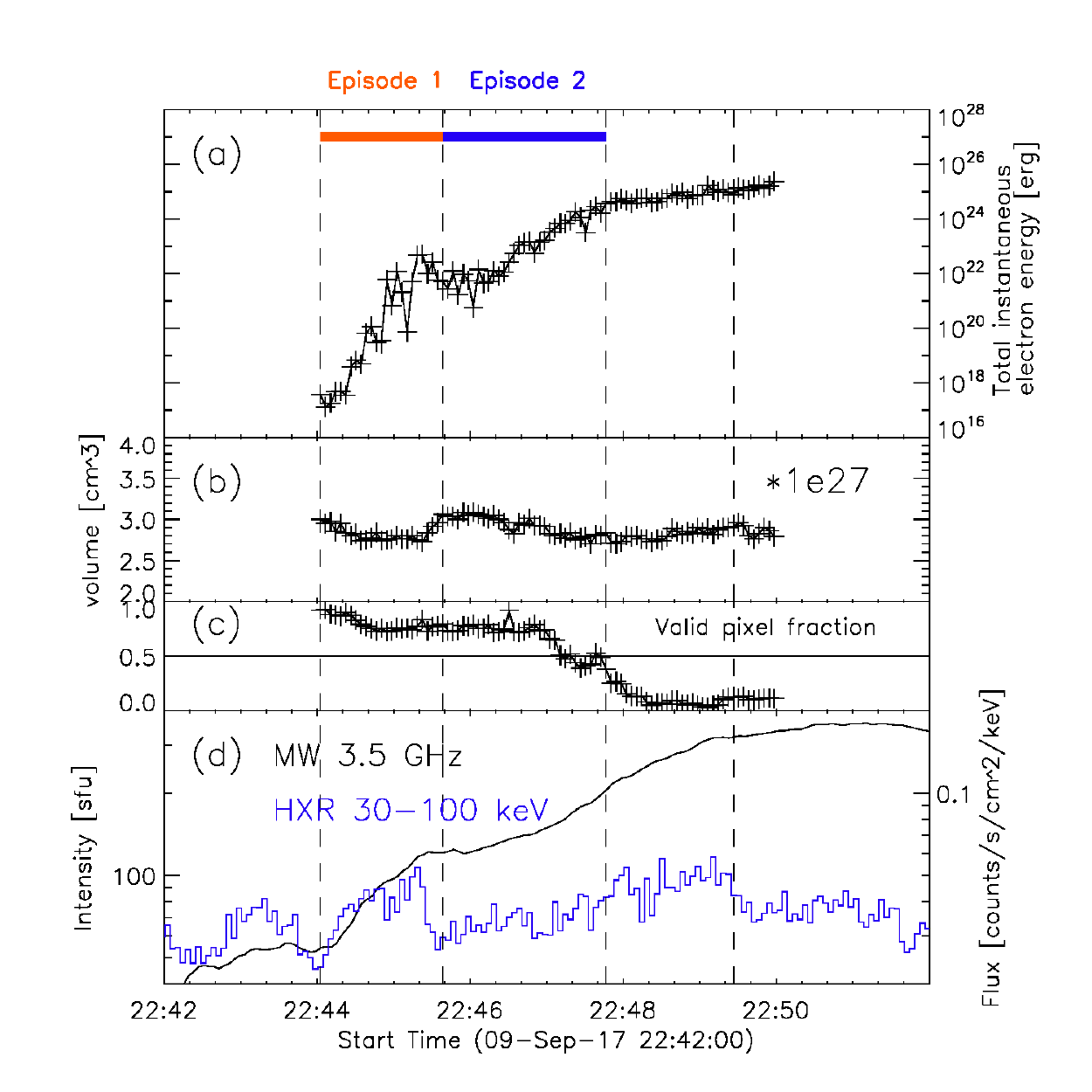}
\caption{Evolution of parameters within the 50\% contour of the observed MW 3.4 GHz source (red contours in Figure \ref{fig:paramovie}). (a) The total instantaneous energy of the nonthermal electrons with energies $>70$ keV. (b) The total volume. (c) The percentage of the total number of pixels in the source that are well-fit and hence selected for analysis. (d) The same lightcurves from Figure \ref{fig:paratrend} (g) for reference. The dashed lines mark times $t_1$--$t_4$.}\label{fig:e_prof5}
\end{figure}

\begin{figure}[ht!]
\includegraphics[scale=1.]{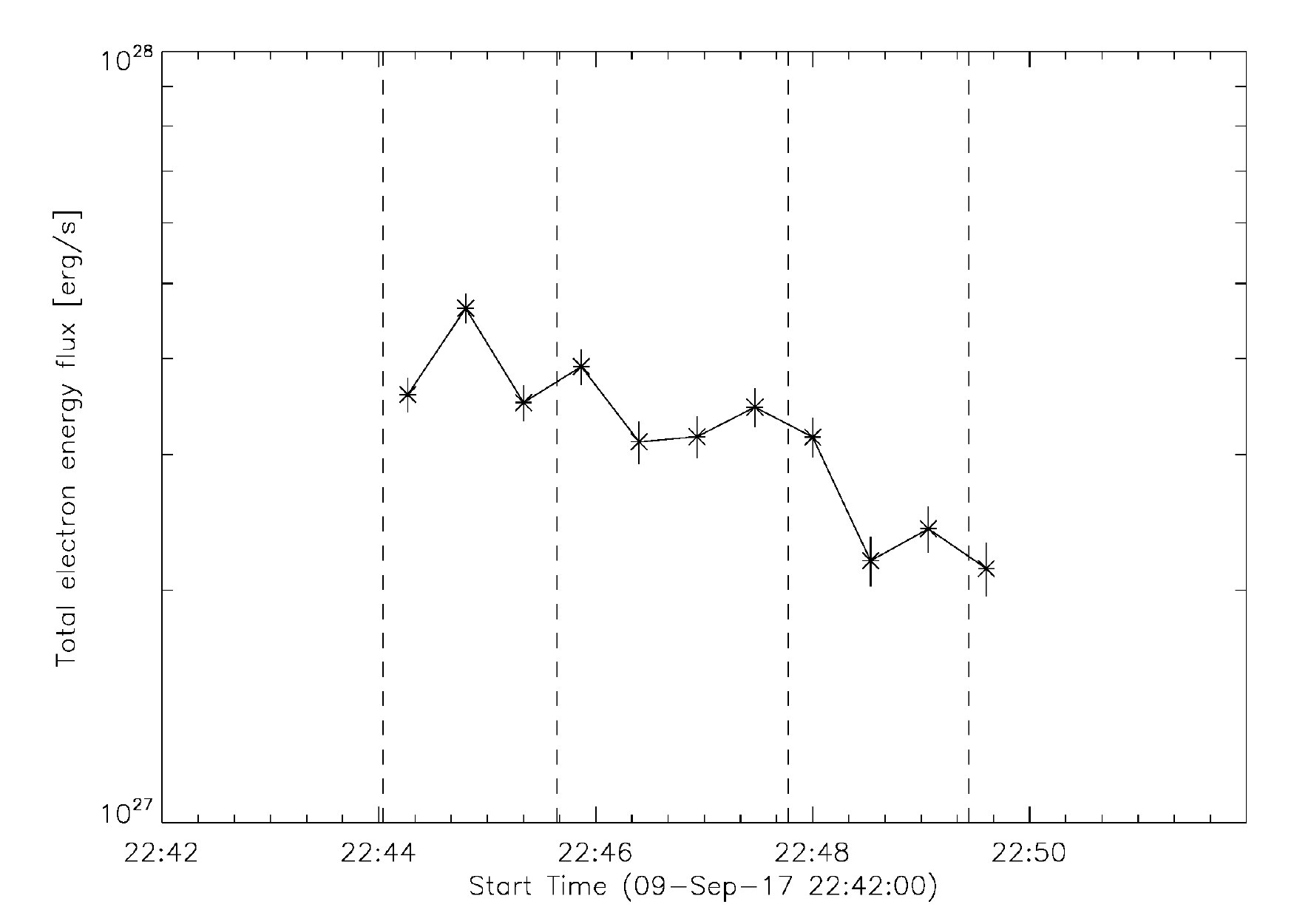}
\caption{The evolution of the total energy flux of HXR-emitting electrons in the chromosphere, using the result of the analysis from Section \ref{sec:analysis_hxr}.}\label{fig:hxren}
\end{figure}


\end{document}